\documentclass[aps,amsfonts,amsmath,prd,preprint,nofootinbib]{revtex4}
\usepackage{amssymb}
\usepackage{amsmath}
\usepackage{graphicx}
\def\beq{\begin{equation}}
\def\eeq{\end{equation}}
\def\bea{\begin{eqnarray}}
\def\eea{\end{eqnarray}}
\def\({\left(}
\def\){\right)}
\def\[{\left[}
\def\]{\right]}

\newcommand{\bel}[1] {\begin{equation}\label{#1}}
\newcommand{\beal}[1] {\begin{eqnarray}\label{#1}}

\begin{document}

\title{Vacuum statistics and stability in axionic landscapes}

\author{Ali Masoumi and Alexander Vilenkin}

\address{Institute of Cosmology, Department of Physics and Astronomy, Tufts University, Medford, MA 02155, USA}

\begin{abstract}

We investigate vacuum statistics and stability in random axionic landscapes.  For this purpose we developed an algorithm for a quick evaluation of the tunneling action, which in most cases is accurate within 10\%.  We find that stability of a vacuum is strongly correlated with its energy density, with lifetime rapidly growing as the energy density is decreased.  
 On the other hand, the probability $P(B)$ for a vacuum to have a tunneling action $B$ greater than a given value declines as a slow power law in $B$.  This is in sharp contrast with the studies of random quartic potentials, which found a fast exponential decline of $P(B)$.  Our results suggest that the total number of relatively stable vacua (say, with $B> 100$) grows exponentially with the number of fields $N$ and can get extremely large for $N\gtrsim 100$.  The problem with this kind of model is that the stable vacua are concentrated near the absolute minimum of the potential, so the observed value of the cosmological constant cannot be explained without fine-tuning.  To address this difficulty, we consider a modification of the model, where the axions acquire a quadratic mass term, due to their mixing with 4-form fields.  This results in a larger landscape with a much broader distribution of vacuum energies.  The number of relatively stable vacua in such models can still be extremely large.

\end{abstract}

\maketitle

\section{Introduction}

The theory of inflation, which has been the leading cosmological paradigm over the last three decades, has led to a major change in our global view of the universe.  According to the new worldview, much of the volume in the universe is still in the state of explosive inflationary expansion.  We live in a "bubble universe", where inflation has ended, but it will never end in the entire space.  The total volume of inflating regions continues to grow, and other bubbles with diverse properties are constantly being formed.  (For a review of the multiverse cosmology, see, e.g., Refs.~\cite{Guth2,Linde2}.)

The dynamics of the inflating multiverse can be described by a rate equation \cite{GSVW}, which includes the energy densities and decay rates of different vacua in the landscape as free parameters.  This dynamics has been studied in some simple models \cite{GSVW,SV}, e.g., in the Bousso-Polchinski \cite{BP} landscape. 
The landscape of string theory and of `realistic' higher-dimensional
models is likely to be much more complicated.  One can hope to gain
some insight into the qualitative features of eternal inflation in
such a landscape by studying vacuum statistics in random potentials.\footnote{Validity of random potentials as models of string landscape has been questioned in Ref. \cite{Dine3}.}
There has been much recent work on vacuum statistics in multi-field
landscapes, e.g., random Fourier \cite{Tegmark,Frazer}, random Gaussian \cite{AazamiEasther,  DeanMajumdar,BrayDean,Douglas,Bachlechner,Battefeld} landscapes and axionic  landscape models \cite{Wang,Higaki, Higaki2, Bachlechner3} , with or without supersymmetry.  Here we shall focus on the non-supersymmetric case.  

A stationary point of the potential, $\partial V/\partial \phi_i = 0$, can be
characterized by its Hessian matrix ${\cal H}_{ij}(\phi) = \partial^2
V/\partial\phi_i \partial\phi_j$.  It is a real, symmetric $N\times N$
matrix, where $N$ is the number of fields in the landscape.  In order
for a stationary point to be a minimum, all eigenvalues of this matrix must
be positive.  For large values of $N$, the probability for this to
happen in a random matrix is extremely small, $P(N) \sim \exp(-\beta
N^2)$, with $\beta\approx 0.27$ \cite{AazamiEasther,DeanMajumdar}.
With $N\gtrsim 100$, this seems to suggest that the landscape contains almost no
metastable vacua. 

It turns out, however, that vacuum statistics in random potentials that are bounded from above and below is
not accurately captured by the random Hessian matrix model.   The
eigenvalues of ${\cal H}_{ij}(\phi)$ are correlated with the potential
$V(\phi)$, and this significantly changes the statistics
\cite{BrayDean,Bachlechner,Battefeld}.  One finds that the probability of finding  a
minimum among stationary points grows towards smaller values of $V$, and that nearly all
stationary points are minima below a certain critical value $V_c$.  The fraction
of minima among the stationary points of the potential is \cite{Bachlechner} $P(N) \propto
\exp(-cN)$ with $c\sim 1$.  The total number of stationary points is expected to
scale in the same way with $c\sim {\rm few}$, and thus the landscape
may have a large number of minima even for $N\gg 1$. 

Another potential problem with a large landscape was highlighted in a
recent paper by Greene et al \cite{Masoumi} (see also \cite{Paban} for a somewhat different result), who argued that vacua in
models with a large number of fields tend to be extremely unstable.\footnote{Concern for metastability of 
vacua in string theory landscape was first raised by Dine et.al \cite{Dine1, Dine2}. }
They approximated the potential near its local minimum by a fourth-order
polynomial with random coefficients and estimated the semiclassical
tunneling rate, approximating the tunneling path by a straight line
leading from the minimum to the lowest saddle point.  The instability
that they found is rather worrisome: as the number of fields $N$ is
increased, the fraction of vacua with sufficiently long lifetimes
decreases  much faster than an exponential.  For $N\gtrsim 100$, this fraction is so
small that the number of such vacua may not be sufficient for the
anthropic explanation of the cosmological constant (even if the peak of the vacuum distribution is "uplifted" to positive energy density).  One of the key assumptions in this analysis is that the set of minima in the landscape is well
represented by an ensemble of polynomials with random coefficients.
It is conceivable, however, that in a bounded potential the decay rate is correlated with
the vacuum energy, with lower-energy vacua having greater stability.

In the present paper, we shall investigate the vacuum statistics and stability in a cosine landscape, defined by the potential
%We use a Fourier landscape model \cite{Tegmark,Frazer,Battefeld,Higaki} 
\beq
V(\phi) = V_0 + \frac{1}{\sqrt{N_c}} \sum_{i=1}^{N_c} A_i \cos \left( \sum_{j=1}^N n_{ij}\phi_j +\alpha_i \right),
\label{landscape}
\eeq
with coefficients $A_i$, phases $\alpha_i$ and integers $n_{ij}$  
%The main
%difference is that \cite{Tegmark,Frazer,Battefeld} included all 
%with Fourier modes 
%with some IR and UV cutoffs, while we include a smaller set of modes
%with $n_{ij}$ 
randomly selected from a suitable distribution.  Without loss of generality we can choose the coefficients $A_i$ to be non-negative, $A_i\geq 0$.  The potential (\ref{landscape}) is obviously bounded from above and below.  
Potentials of this form can be expected, e.g., in the axion sector of string theory.  We note that the cosine landscape (\ref{landscape}) is different from the Gaussian Fourier landscapes discussed in Refs.~\cite{Tegmark,Frazer}, which include all Fourier modes with some IR and UV cutoffs, while we include a fixed number $N_c$ of cosines with randomly selected linear combinations of the fields $\phi_j$ in the cosine arguments.  

In the next section we specify the details of our cosine landscape, and in Section III, we estimate the total number of vacua in the landscape and find their energy distribution.  One of the goals here is to see how the results compare with random Gaussian models, and thereby to what extent such models can represent a generic random potential.    In Section IV, we study the decay rate of the vacua, which we characterize by the tunneling action $B$, and find the probability distribution for $B$.  

A potential problem with the axionic and similar models is that relatively stable vacua tend to be concentrated near the absolute minimum of the potential, so the observed value of the vacuum energy density cannot be explained without fine-tuning. To address this difficulty, we consider a modification of the model, where the axions acquire a quadratic mass term, due to their mixing with 4-form fields.  This model, its vacuum statistics and stability are analyzed in Section V.  Our conclusions are briefly summarized in Section VI.

\section{The model}

We shall consider an ensemble of random cosine landscape models with potentials $V(\phi)$ of the form (\ref{landscape}).  We choose the coefficients $A_i$ from a Gaussian distribution with zero mean and standard deviation 
\beq
\Delta_A = \lambda M^4.  
\label{DeltaA}
\eeq
Here, $M$ is the characteristic energy scale and $\lambda$ is a dimensionless coupling constant.  The fields $\phi_i$ are also dimensionless.  In our simulations we used units where $M=1$ and set $\lambda=0.1$.
Note that the choice of $\lambda$ affects only the overall normalization of the potential.  It has no effect on vacuum statistics and results in a simple rescaling of the tunneling action. 
The potential (\ref{landscape}) has a shift symmetry
\beq
\phi_j \to \phi_j + 2\pi n_j,
\eeq
where $n_j$ is an integer.  We shall assume that the fields $\phi_j$ take values in the range
$0\leq \phi_j < 2\pi$.

%$2^{1/2}N_c^{-1/2} M^4$, where $N_c$ is the number of cosine terms in the potential (\ref{landscape}) and $M$ is the characteristic energy scale of the landscape.  The factor $N_c^{-1/2}$ is introduced in order to ensure that the mean square variation of the potential is $\langle (V-V_0)^2 \rangle^{1/2} = M^4$.  (Here and below, angular brackets indicate averaging over the ensemble.)

\begin{figure}[htbp] %  figure placement: here, top, bottom, or page
   \centering
     \includegraphics[width=3in]{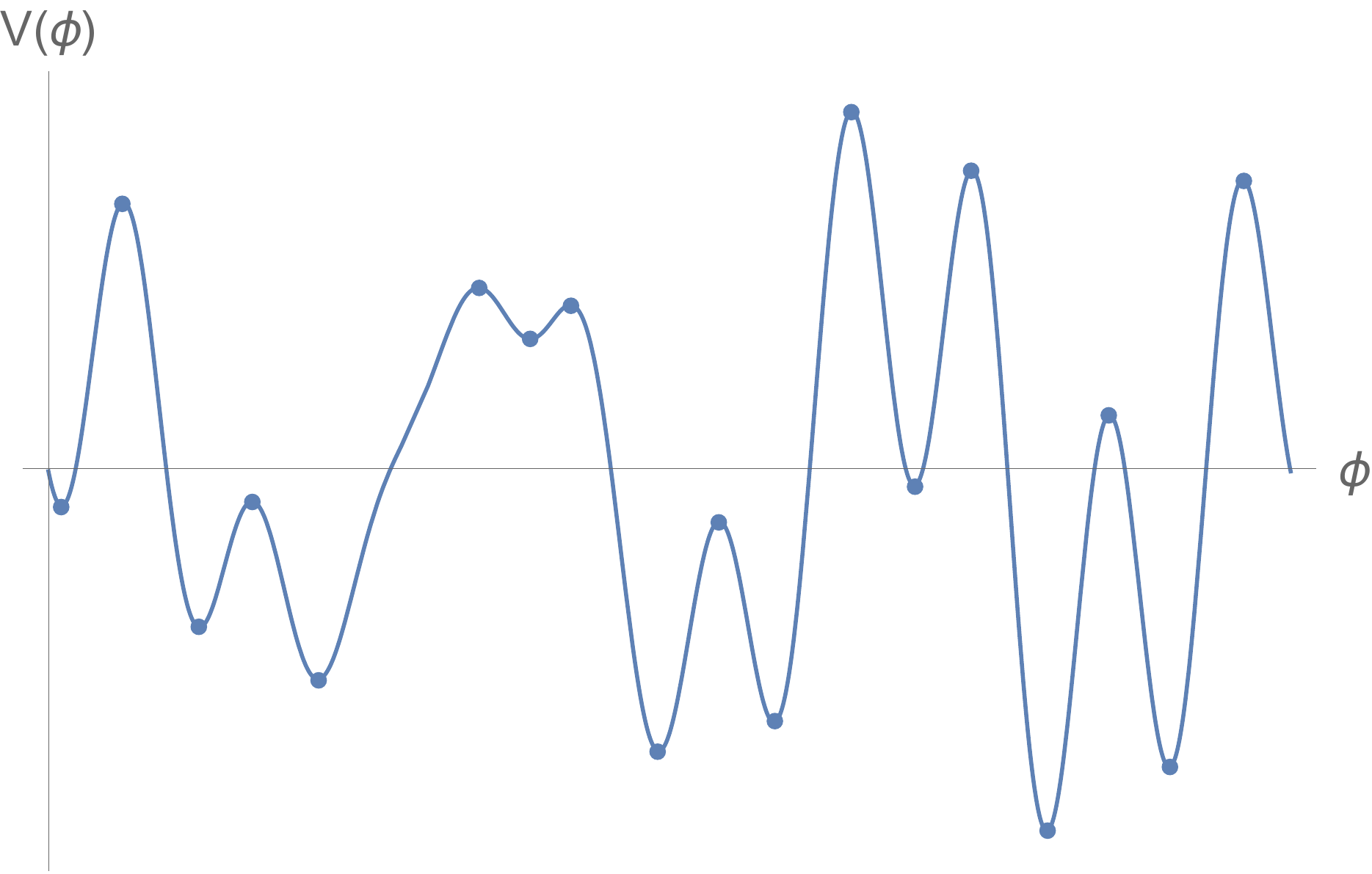} 
    \includegraphics[width=3in]{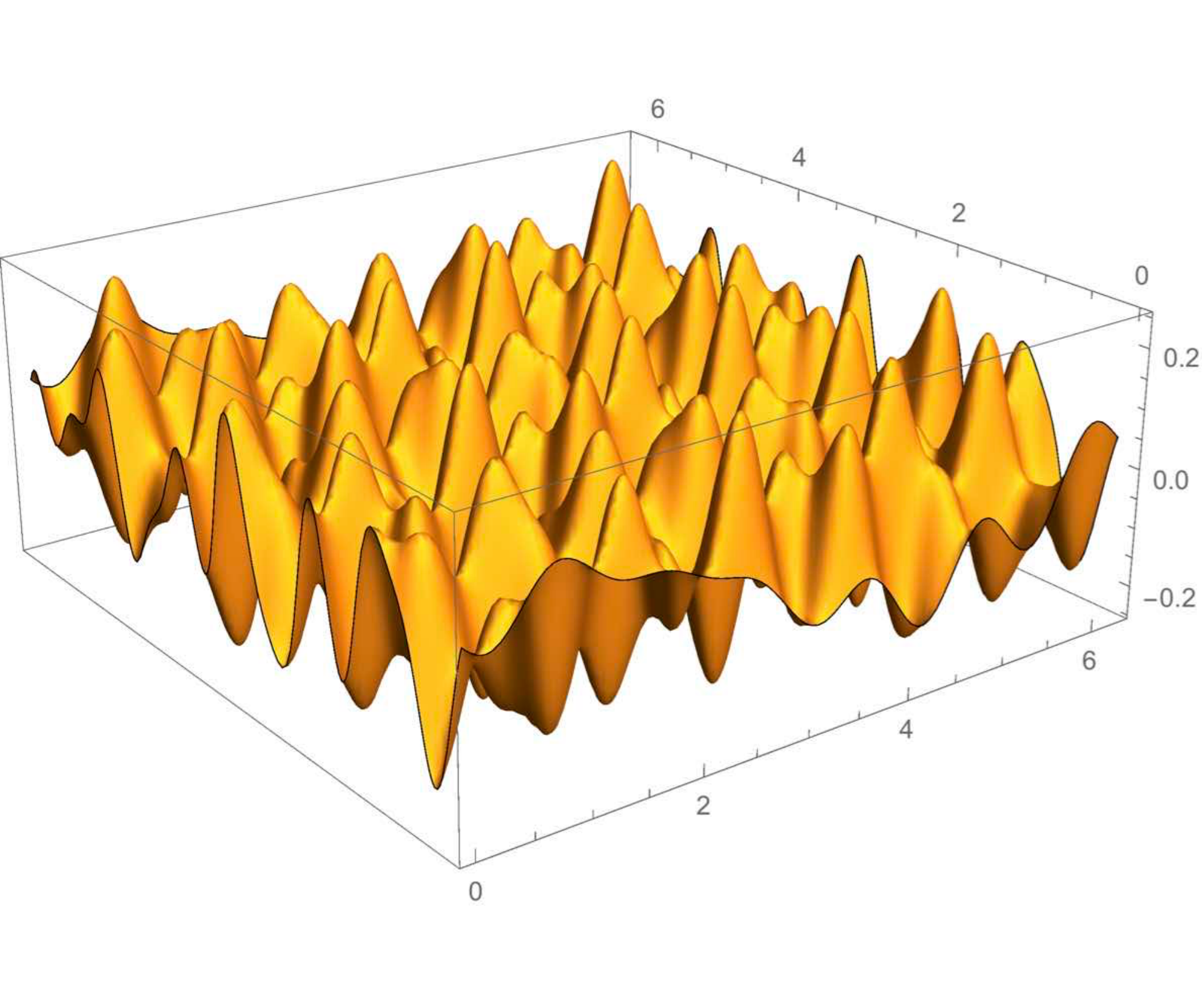}
   \caption{Two realization of the potential described in \eqref{landscape} for one and two fields. }
   \label{fig:twoPotentials}
\end{figure}

The phases $\alpha_j$ in Eq.~(\ref{landscape}) are chosen at random in the range $[0, 2\pi]$, and the integers $n_{ij}$ are chosen from a uniform distribution in the range $[-n_{max}, n_{max}]$.  The parameter $n_{max}$ determines the characteristic distance $\delta\phi$ in the field space between the stationary points of $V(\phi)$,
\beq
\delta\phi_j \sim 2\pi / n_{max}.
\eeq
Unless explicitely stated otherwise, we used the values $N_c=30$, $n_{max}=10$ in all our simulations. 
Typical realizations of the potential (\ref{landscape}) for $N=1$ and 2 fields are shown in Fig. \ref{fig:twoPotentials}.

For given values of $n_{max}$ and $N$, the total possible number of independent terms in the potential (\ref{landscape}) is $(2n_{max}+1)^N -1$.  For $N>1$ and with our standard values of $n_{max}=10$ and $N_c=30$, this is much greater than the actual number of cosine terms $N_c$.  For $N=1$, on the other hand, $N_c$ is greater than the number of independent terms, so we can expect (almost) all possible Fourier modes to be represented.

In the context of string theory, the cosine terms in the axion potential (\ref{landscape}) are generated by instantons, with the coefficients $A_i \propto \exp(-S_i)$, where $S_i$ is the corresponding instanton action.  One can expect therefore that the magnitudes of $A_i$ are uniformly distributed on a logarithmic scale \cite{axiverse}.  Inclusion of subleading contributions in (\ref{landscape}) may result in a band structure of the vacuum energy spectrum \cite{Bachlechner3}.  Here we disregard such contributions, so our model can be regarded as describing only the leading part of the axion landscape.  More generally, it can be regarded as representing a generic landscape with a potential bounded from above and below.

A generic Lagrangian for our cosine landscape has the form
\beq
{\cal L} = \frac{1}{2} \sum_{i,j=1}^N K_{ij}\partial_\mu \phi_i \partial^\mu \phi_j - V(\phi).
\eeq
The vacuum energy distribution depends only on the potential $V(\phi)$, but the kinetic terms will also be important for the analysis of vacuum stability.  The matrix $K_{ij}$ plays the role of the metric in the $N$-dimensional field space.  For simplicity, we shall assume that $K_{ij}$ is proportional to a unit matrix, 
\beq
K_{ij} = f^2 \delta_{ij},
\label{Kij}
\eeq
where $f$ is a constant parameter with the dimension of energy.  Furthermore, we shall assume that the two energy scales of the model are the same: $f=M=1$.  Otherwise, any additional factor can be absorbed into a redefinition of $\phi_i$, resulting in a rescaling of the tunneling action.  Our analysis can be easily extended to a more general form of $K_{ij}$.

An  important property of potentials of the form (\ref{landscape}) is that for $N_c\le N$ there are $(N-N_c)$ flat directions and the potential has a unique value of the vacuum energy.  The easiest way to see this is to choose linear combinations $\Phi_i =\sum_{j=1}^{N} n_{ij} \phi_j+ \alpha_i$ for $i=1, \ldots, N_c$. For the rest of the $N-N_c$ directions we choose $\Phi_i$ to be orthogonal to these linear combinations. Then the potential in \eqref{landscape} simplifies to 
 \begin{equation}\label{eq:pot9}
	V= V_0 + N_c^{-1/2} \sum_{i=1}^{N_c} A_i\cos\left(\Phi_i \right)~.
\end{equation}
The local minima are at $\Phi_i=\pi$ with vacuum energy  $V_0 - N_c^{-1/2} \sum_{i=1}^{N_c} A_i$. As we increase $N_c$ beyond $N$ we see a distribution for different values of vacuum energy and the distribution gets wider for larger $N_c$. 

%\subsection{Case of sum of independent potentials}

We note also that the spectrum of vacuum energies can be easily characterized in
the simple case where the potential can be represented as 
\bel{indepSum}
	V(\phi_1, \ldots, \phi_N)= \sum_{i=1}^N V_i(\phi_i)~.
\eeq
In order to get a vacuum, we need to have a minimum in all field directions. But  the one-dimensional potentials $V_i$ have equal numbers of maxima and minima, and therefore the chance that a given stationary point is a local minimum is exactly $2^{-N}$.
If the values of vacuum energy of $V_{i}$'s are drawn from the same distribution, we can immediately infer the spectrum of the theory for $N\gg 1$ from the central limit theorem. If the spectrum of vacuum energies of each $V_i$ has a distribution with standard deviation $\sigma$ and average $ \mu$, the distribution of vacuum energies of the theory is given by 
\bel{vacUDist}
	P(V_{\rm vac})= \frac{1}{\sigma \sqrt{2\pi N}} \exp\[-\frac{(V_{\rm vac} - \mu N)^2}{2 \sigma^2 N} \]~.
\eeq
For landscapes without an offset (i.e. $V_0=0$ in \eqref{landscape}) we expect to have $\mu <0$. Therefore, in the large N limit almost all the vacua would correspond to AdS spaces.

\section{Vacuum statistics}\label{sec:Numeric}

%To generate random potentials introduced in \eqref{landscape} that, we choose $A_i$'s from a (absolute value)  Gaussian distribution of zero mean and standard deviation $\sigma_{A}$, and $n_{ij}$'s from a flat distribution of integers in the range $[-n_{\rm max},n_{\rm max}]$. 
%Because the potentials (\ref{landscape}) are periodic, it suffices to study the statistics in a hypercube  $[0,2\pi]^{N}$.  
Starting from a grid with a lattice spacing $\pi/2n_{max}$ (a quarter of the smallest wavelength), we can expect to find (most of the times) all the critical points of the potential (\ref{landscape}).  However, the required size of the grid grows rapidly with the number of fields and becomes prohibitively large even for modest values of $N$ and $n_{max}$.  We therefore used this method only for $N=1,2,3$  and 4 and used a Monte-Carlo sampling to study the vacuum statistics for larger values of $N$.  The details of our numerical procedure are given in Appendix \ref{sec:findRoot}. 

\begin{figure}[htbp] %  figure placement: here, top, bottom, or page
   \centering
   \includegraphics[width=2.5in]{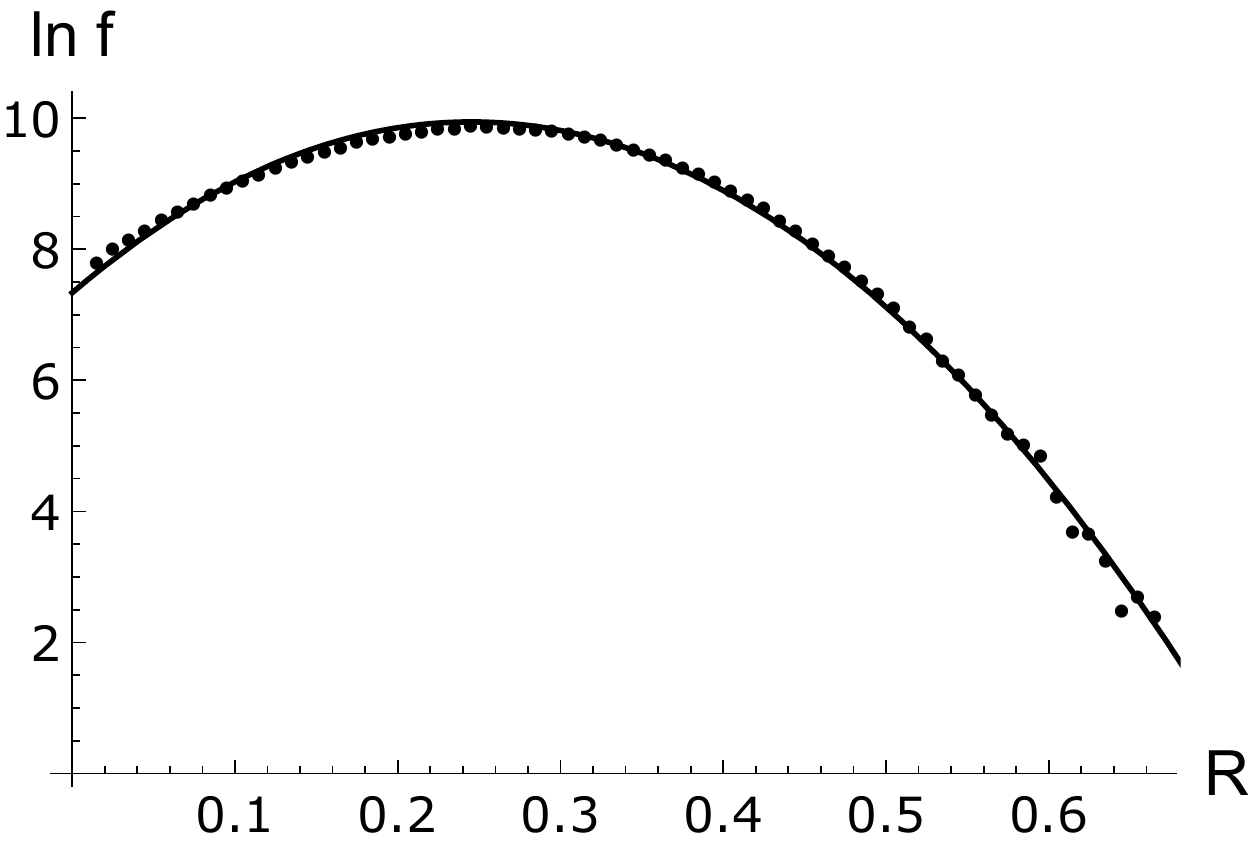} 
    \includegraphics[width=2.5in]{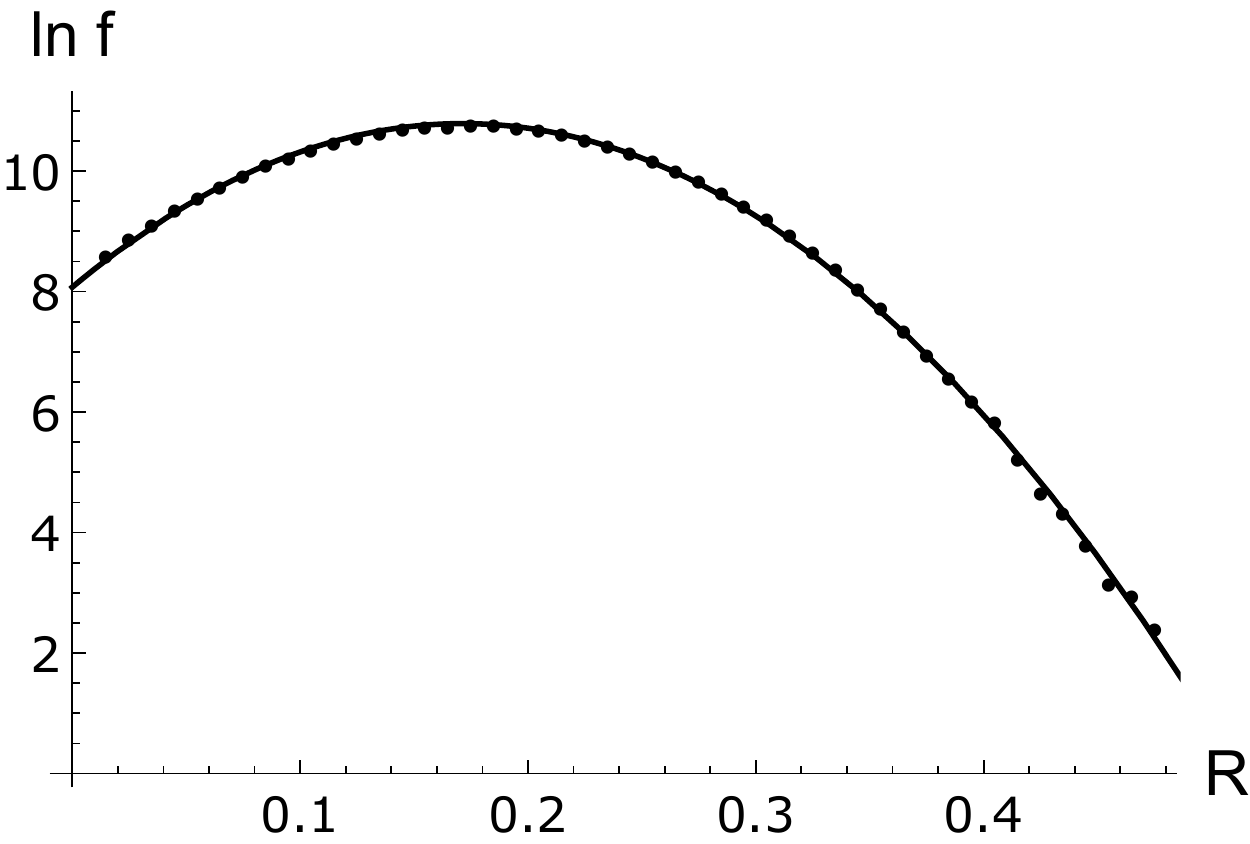} 
    \includegraphics[width=2.5in]{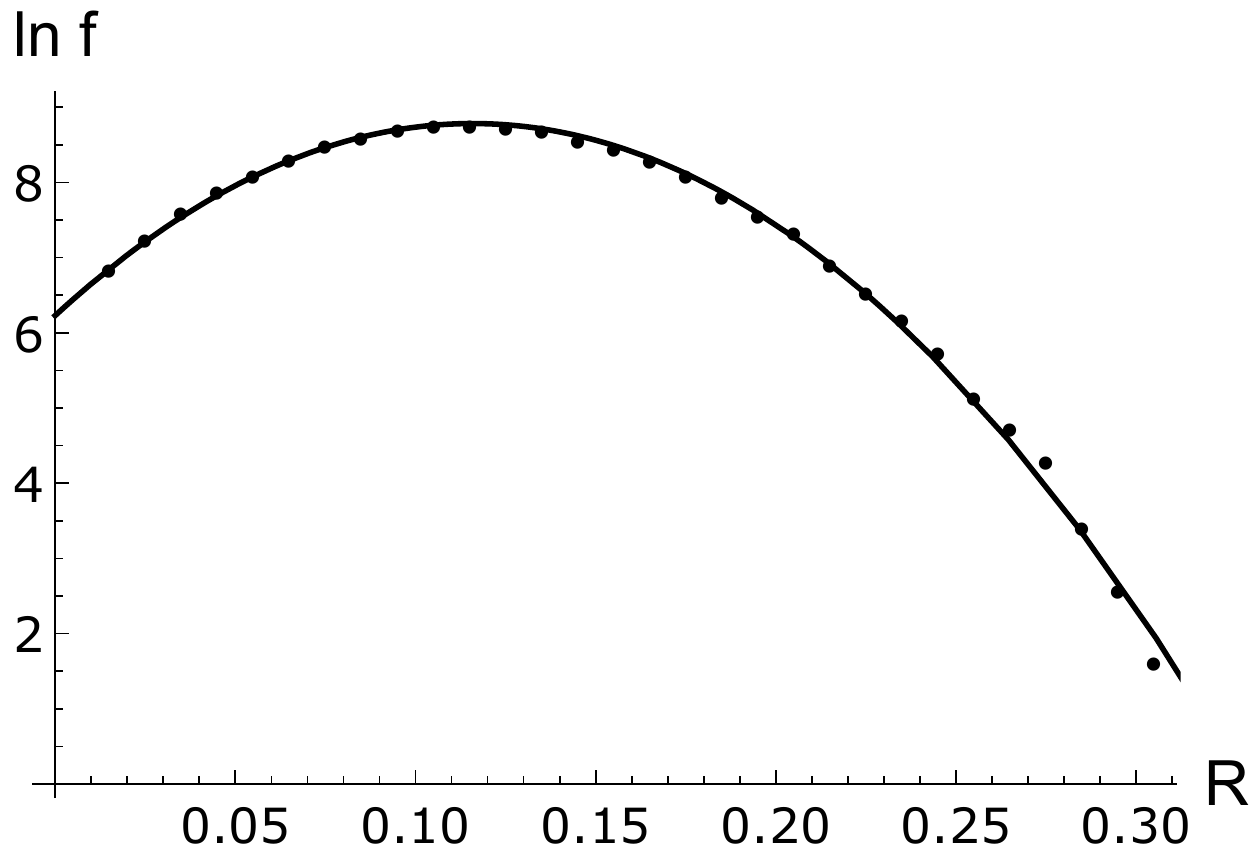} 
    \includegraphics[width=2.5in]{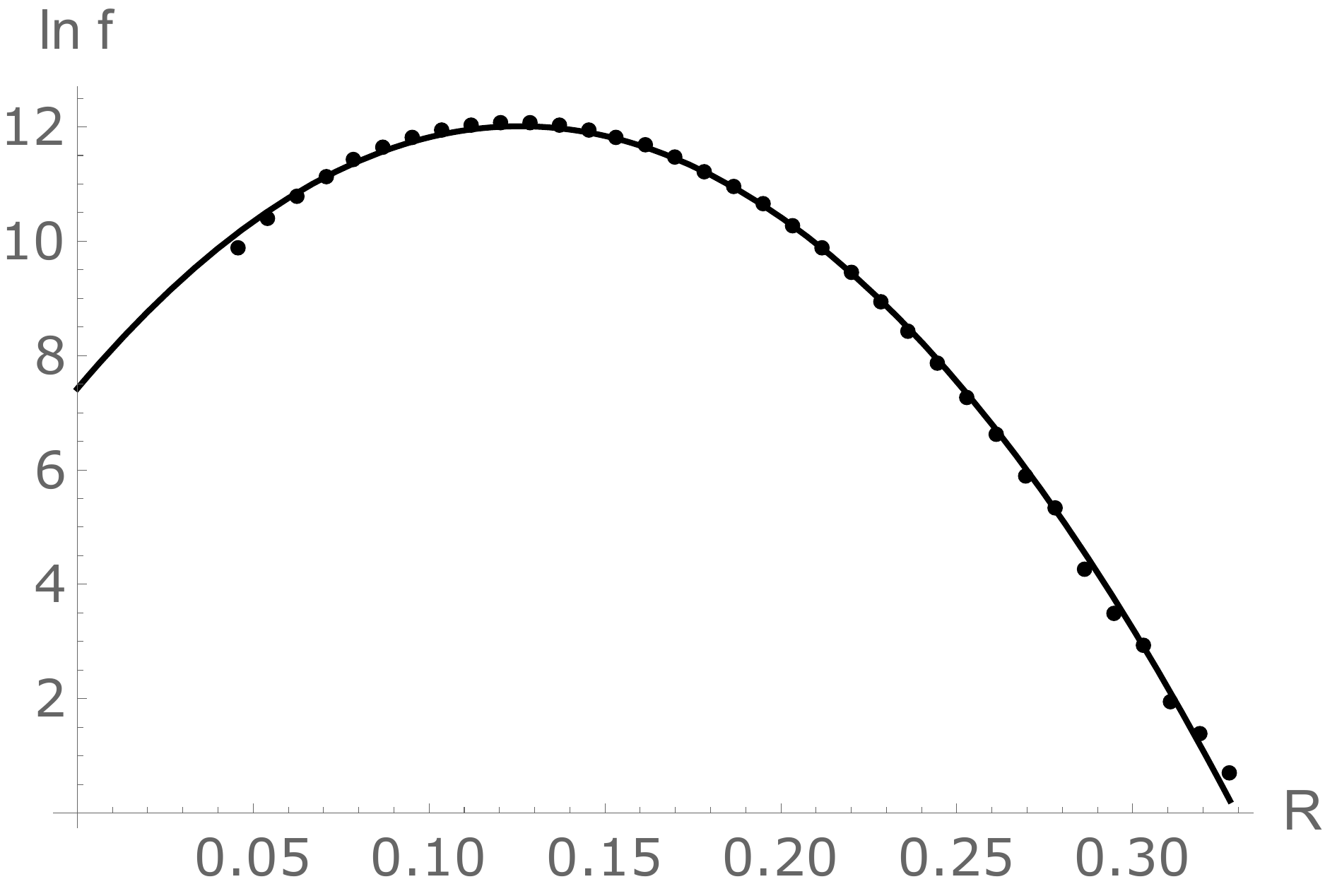} 
   \caption{ Logarithm of the distribution $f(R)$ fitted with a quadratic form for different parameter values.  From top left to bottom right, $N=2,4,7$ and 8.  In all simulations here and below, except where explicitly stated otherwise, we used the values $N_c=30$ and $n_{\rm max}=10$.}
%{Distribution of the log of $R$ for different parameters. In  all of them the $R$'s are almost perfect Gaussians.}
   \label{fig:RUDistOct27}
\end{figure}

Different realizations of the potential, corresponding to different choices of $A_i$ and $n_{ij}$ in (\ref{landscape})
will generally have different ranges of variation.  However, we are more interested in understanding how the vacuum energies are distributed with respect to each other, e.g., whether they clump up near the global minimum or spread uniformly in the available range. To make this aspect of the distribution manifest, we define a quantity 
which tells us where the vacuum in a sample is located with respect to the global minimum and maximum of the potential. If for a given realization of the potential the global minimum and maximum have energies $V_{\rm min}$ and $V_{\rm max}$, then for a minimum of energy $V$ in that sample we define 
\beq
	R = \frac{V-V_{\rm min}}{V_{\rm max}-V_{\rm min}}.
\label{RDef}
\eeq
This quantity can range between 0 and 1. 
%For a typical potential with no {\bf offset}, $V_{\rm min} \approx - V_{\rm max}$. Therefore, $R>1/2$ corresponds to a dS space and $R<1/2$ is for an AdS space. 
For each realization of the potential we found the local minima and calculated their values of $R$.  We divided these values into bins and plotted the frequency of occurrence of different values $f(R)$ vs $R$. The results for several values of $N$ are shown in Fig.\ref{fig:RUDistOct27}, where we also fitted $\ln f(R)$ with a quadratic form.  It is apparent from the Figure that the distributions $f(R)$ are nearly perfect Gaussians,\footnote{We note that the corresponding distributions for the vacuum energy $V$ are not well approximated by Gaussians.}   
\beq
f(R)\propto \exp \left(-\frac{(R-R_{m})^2}{2\sigma_R^2}\right),
\label{Gaussian}
\eeq
at least for values of $N$ between 1 and 10.

\begin{figure}[htbp] %  figure placement: here, top, bottom, or page
   \centering
    \includegraphics[width=2.5in]{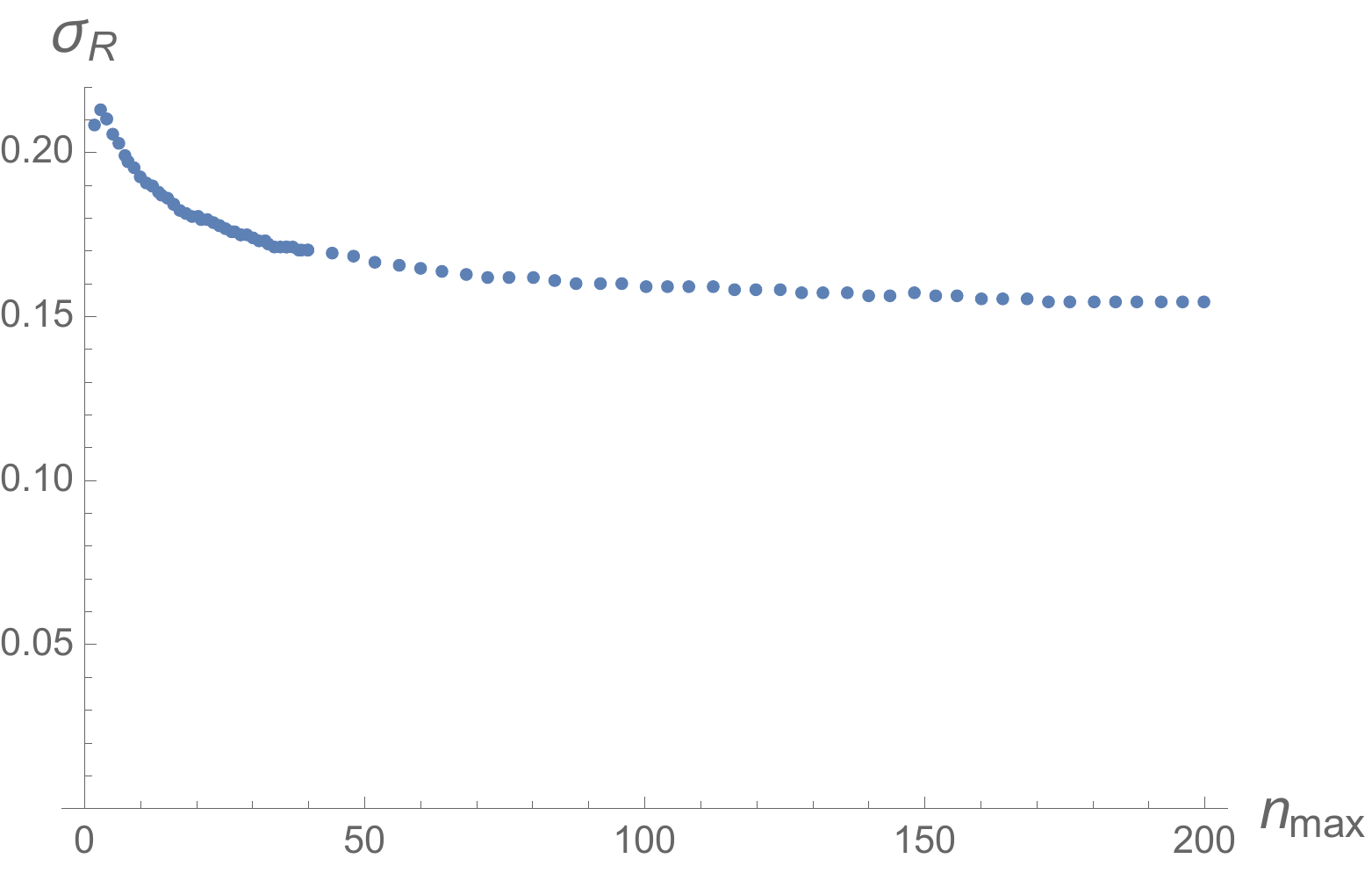} 
    \includegraphics[width=2.5in]{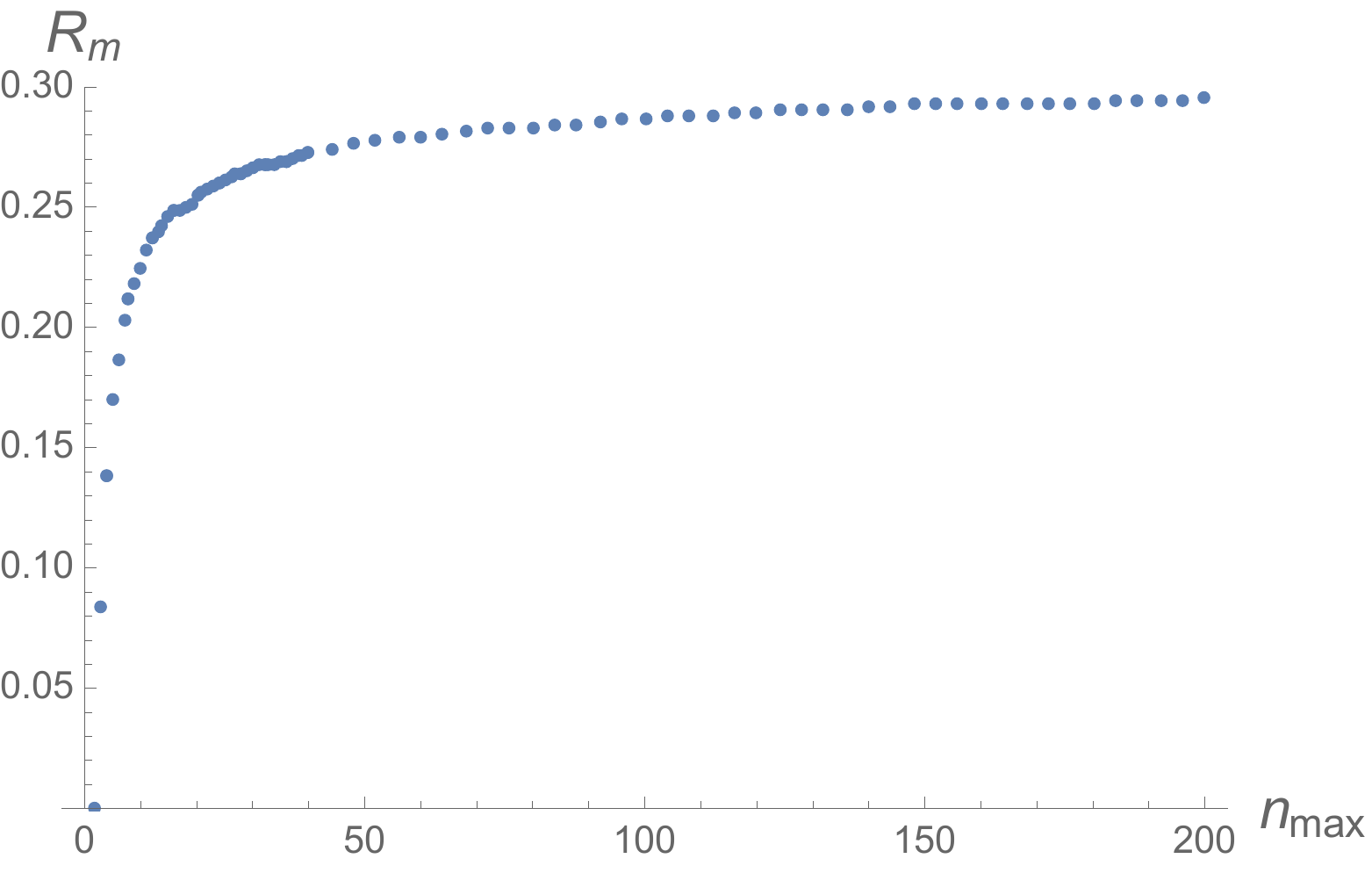} 
   \caption{The standard deviation $\sigma_R$ and the peak value $R_m$ vs $n_{\rm max}$ for one field $(N=1)$ and $N_c=6$.}
   \label{fig:rMedianLargenMax}
\end{figure}

The parameters $R_m$ and $\sigma_R$ characterizing the distribution (\ref{Gaussian}) are rather insensitive to the values of $n_{max}$ and $N_c$, as long as $n_{max}$ and $N_c - N$ are significantly greater than 1.  As an example, we plot in Fig.\ref{fig:rMedianLargenMax} $R_m$ and $\sigma_R$ vs. $n_{max}$ for one field ($N=1$) and $N_c=6$.  We see that the variation of both parameters is relatively small at $n_{max}\gtrsim 10$ and that they approach fixed asymptotic values at large $n_{max}$.  We have verified that for $N=1,2$, $R_m$ and $\sigma_R$ vary by no more than 20\% and 10\%, respectively, as $n_{max}$ and $N_c$ vary in the range $8\leq n_{max} \leq 20$, $6\leq N_c \leq 12$.

\begin{figure}[htbp] %  figure placement: here, top, bottom, or page
   \centering
    \includegraphics[width=3.2in]{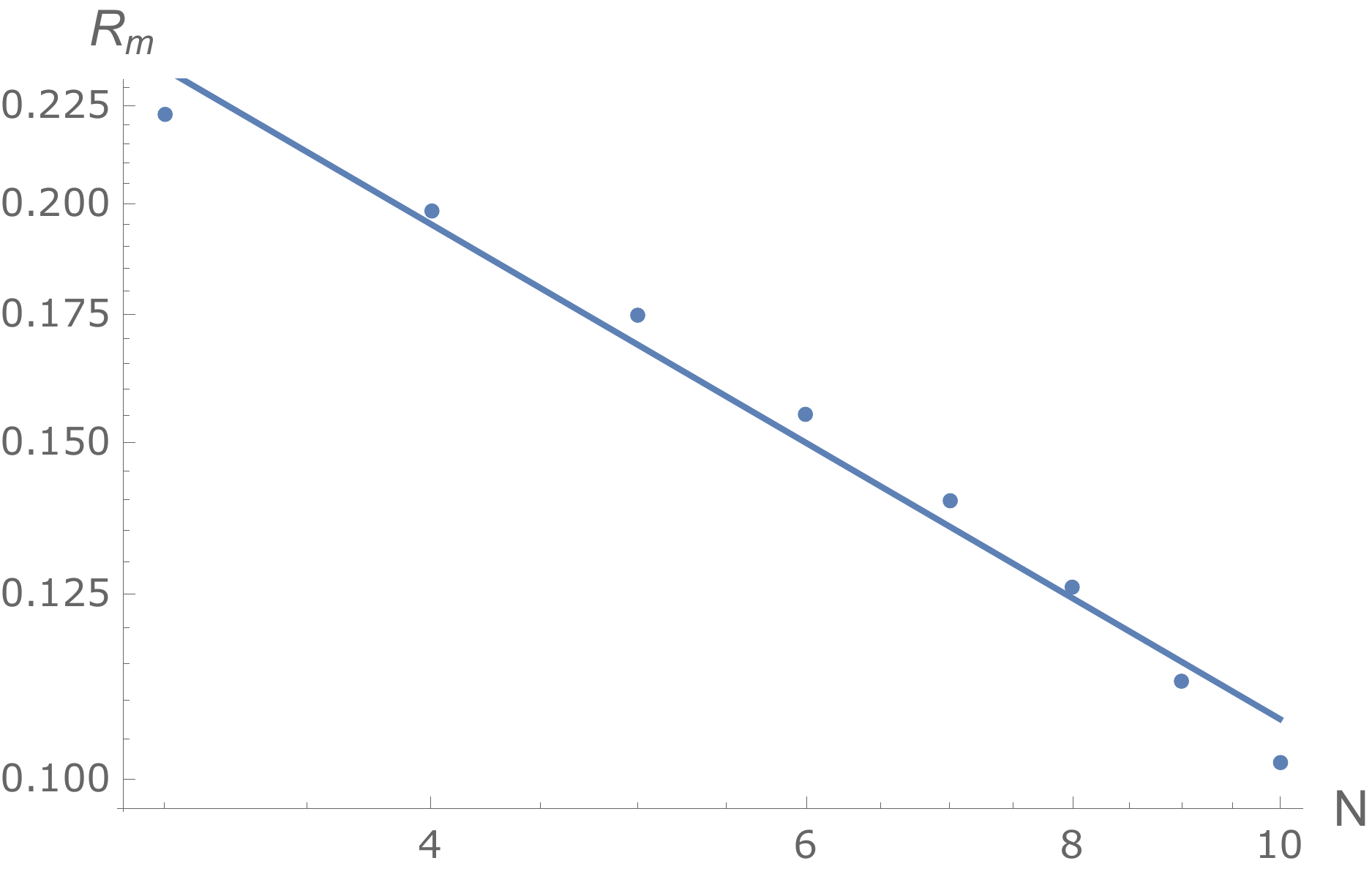}
    \includegraphics[width=3.2in]{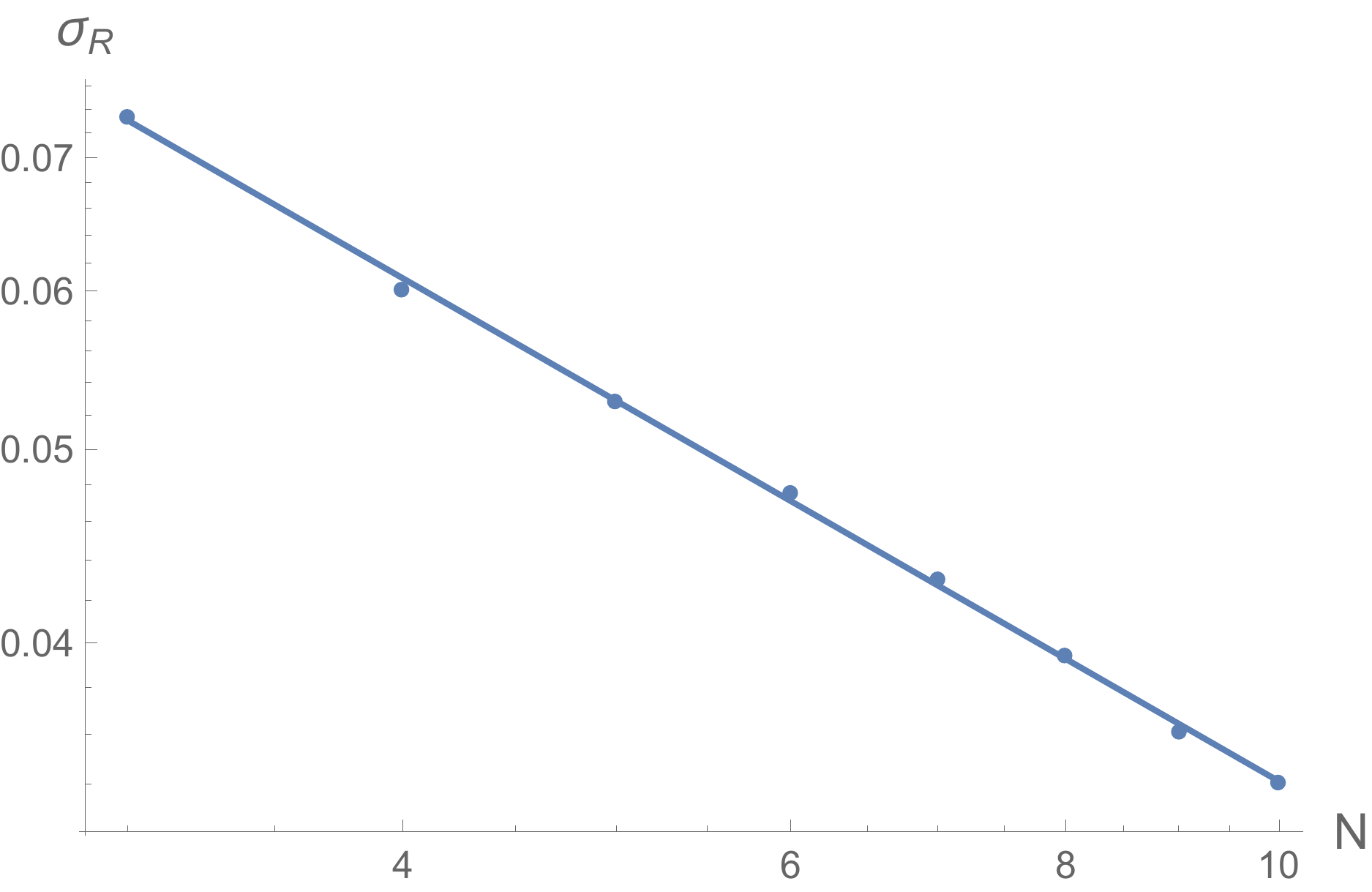}
   \caption{The Gaussian distribution parameters $R_m$ and $\sigma_R$ vs. the number of fields $N$ on a log-log plot.}  
   %Here we used the fixed values $N_c = 30$, $n_{max} = 10$.}
   \label{fig:RVsNf}
\end{figure}

On the other hand, the distributions $f(R)$ do show a significant dependence on the number of fields $N$.  As $N$ grows with other parameters fixed, both $R_m$ and $\sigma_R$ decrease, so the distribution becomes more and more concentrated near the global minimum $R=0$.  The dependence $\sigma_R (N)$ is well fitted by a power law,
\beq
\sigma_R \propto N^{-\alpha}
\label{sigmaN}
\eeq
with $\alpha \approx 0.63$, while the decline of $R_m$ with $N$ is faster than a power-law; see Fig.\ref{fig:RVsNf}.  We note that the dependence (\ref{sigmaN}) cannot extend to arbitrarily large values of $N$.  When $N$ reaches the value $N=N_c$, the distribution degenerates into a delta-function, $f(R)=\delta(R)$, which corresponds to $R_m =\sigma_R =0$.  Hence, the dependence (\ref{sigmaN}) can be expected only for $N_c - N > 1$.

\begin{figure}[htbp] %  figure placement: here, top, bottom, or page
   \centering
    \includegraphics[width=3.2in]{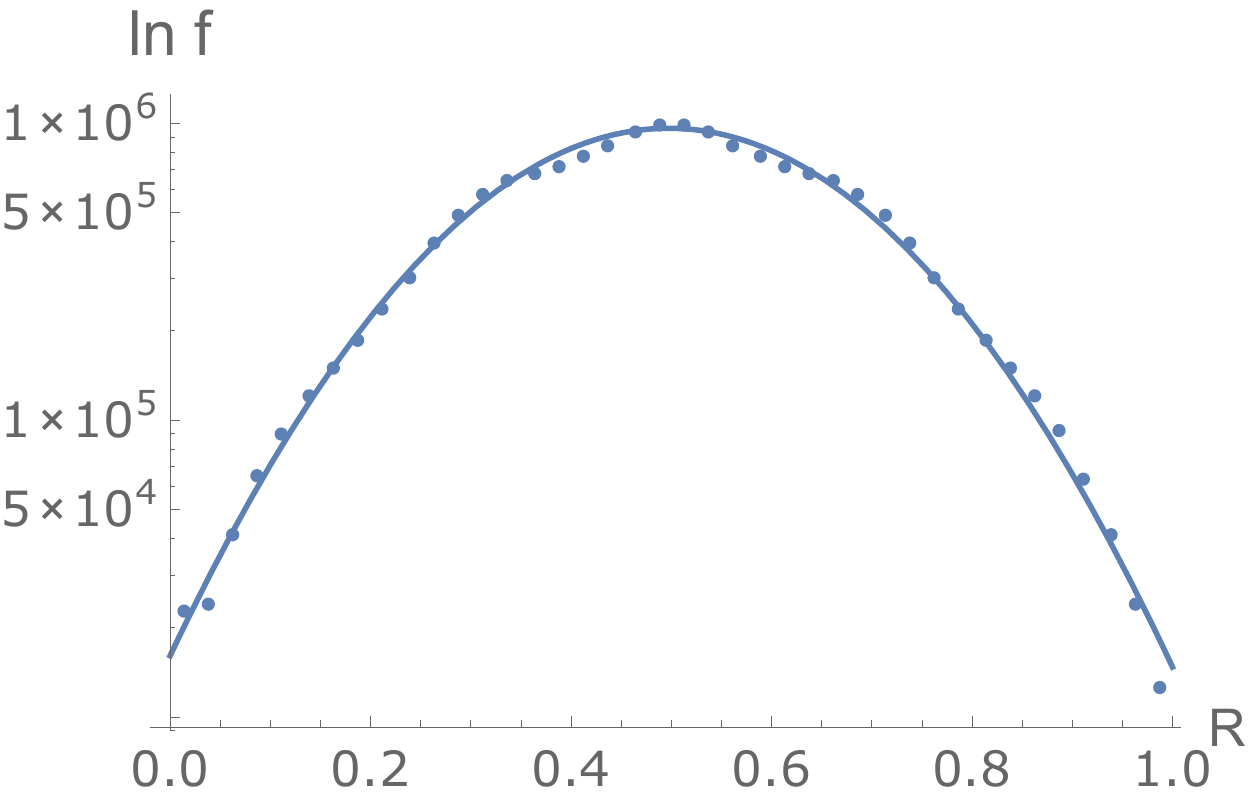} 
    \includegraphics[width=3.2in]{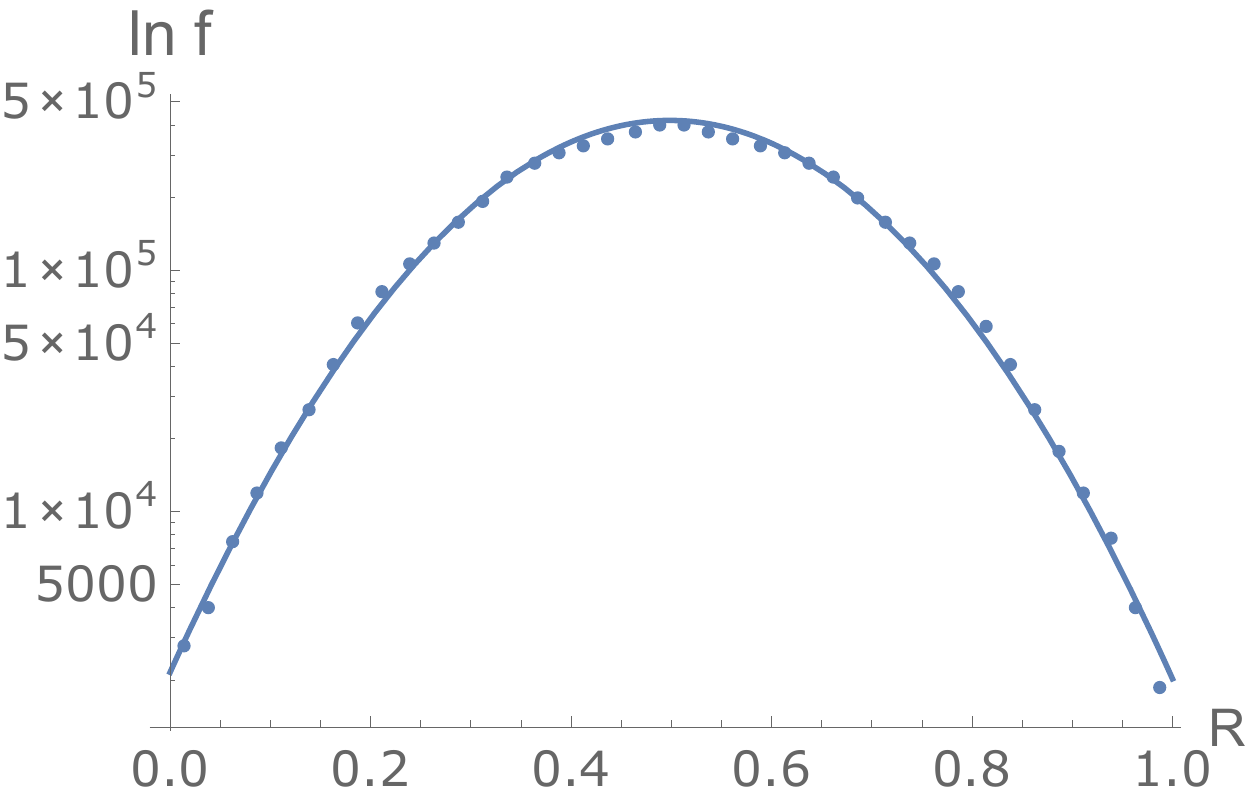}
   \caption{Distribution of stationary points vs $R$ for $N=4$ (left) and 6 (right). This is well fitted by a Gaussian.}   
   \label{fig:StationaryNumVsR}
\end{figure}

To study the statistics of stationary points of the potential, we found all stationary points for $N=1,2,3$ and 4 and sampled large numbers of them for larger $N$. 
%{\bf [Should we say how many realizations we used?]}
The resulting distributions are well approximated by Gaussians peaked at $R=0.5$ with a width scaling as $\sigma_{st} \propto N^{-0.33}$.  The distributions for $N=4$ and 6 are shown in Fig.\ref{fig:StationaryNumVsR}.    

\subsection{Probability of a minimum} \label{subsec:Probability}

An important characteristic of our model is the probability for a stationary point of the potential to be a minimum,
\beq
P_{\rm min}(R) = \langle \frac{{\cal N}_{\rm min}(R)}{{\cal N}_{\rm st}(R)} \rangle,
\eeq
where angular brackets indicate averaging over the ensemble.
Here, ${\cal N}_{\rm min}(R)$ and ${\cal N}_{\rm st}(R)$ are respectively the number of minima and the number of stationary points at a given value of $R$. (These quantities are evaluated within a small interval $\Delta R$, but we expect them to be insensitive to the magnitude of $\Delta R$.)  If one assumes naively that the sign of the $N$ eigenvalues of the Hessian is chosen randomly, then the chance for a given stationary point to be a minimum would be $2^{-N}$ for all values of $R$.  
On the other hand, Figs. \ref{fig:RUDistOct27} and \ref{fig:StationaryNumVsR} demonstrate that ${\cal N}_{\rm min}$ and ${\cal N}_{\rm st}$ do in fact depend on $R$, suggesting that $P_{\rm min}$ should also be strongly $R$-dependent.

%one can expect $P_{\rm min}(R)$ to grow as $R$ decreases; in particular, we should have $P_{\rm min}(R=0) = 1$.

\begin{figure}[htbp] %  figure placement: here, top, bottom, or page
   \centering
    \includegraphics[width=3.2in]{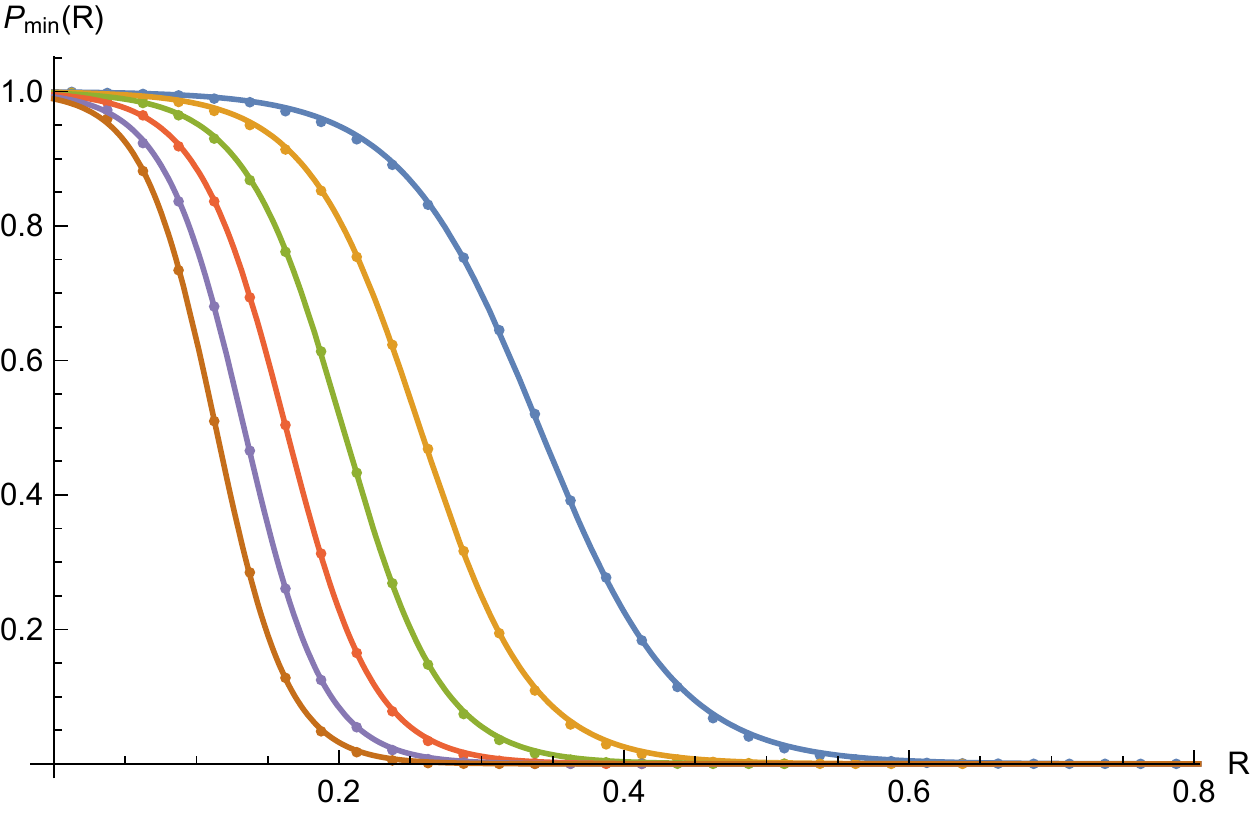}
    \includegraphics[width=3.2in]{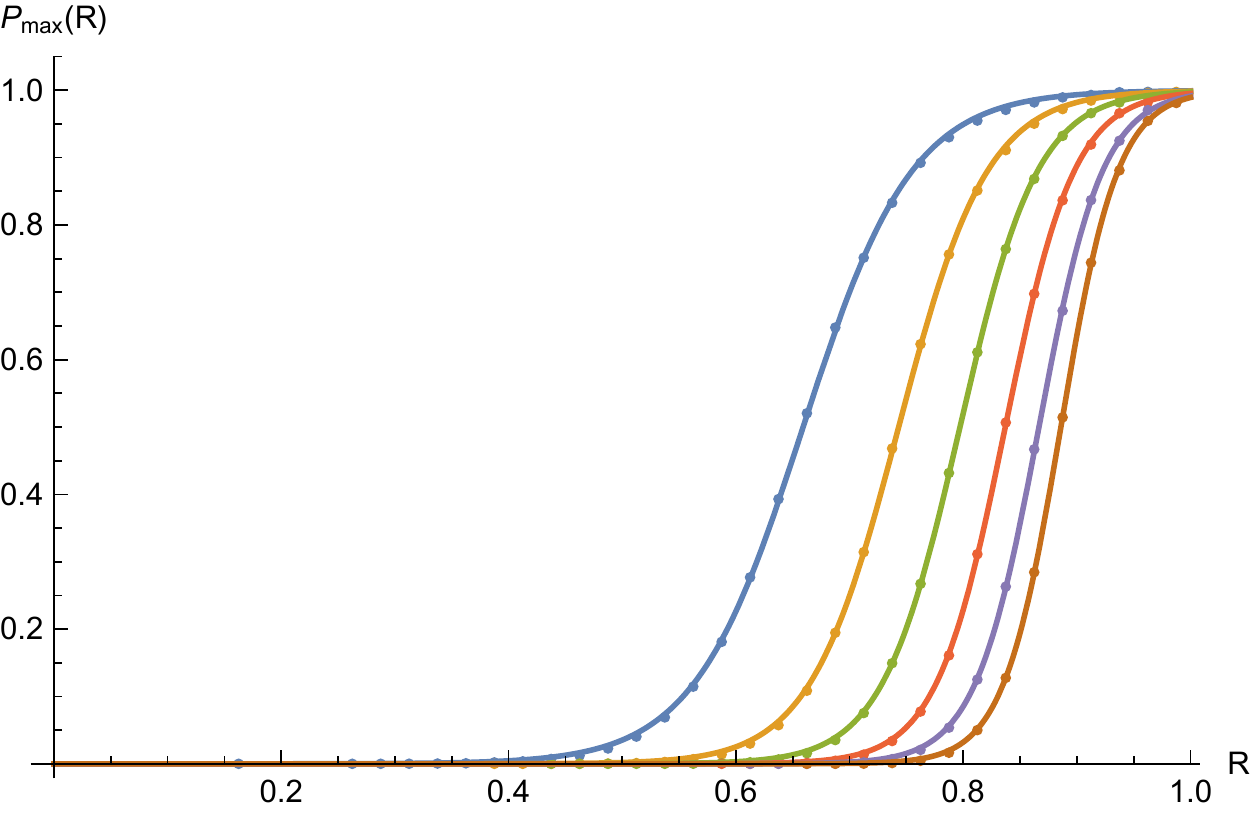}
    \includegraphics[width=3.2in]{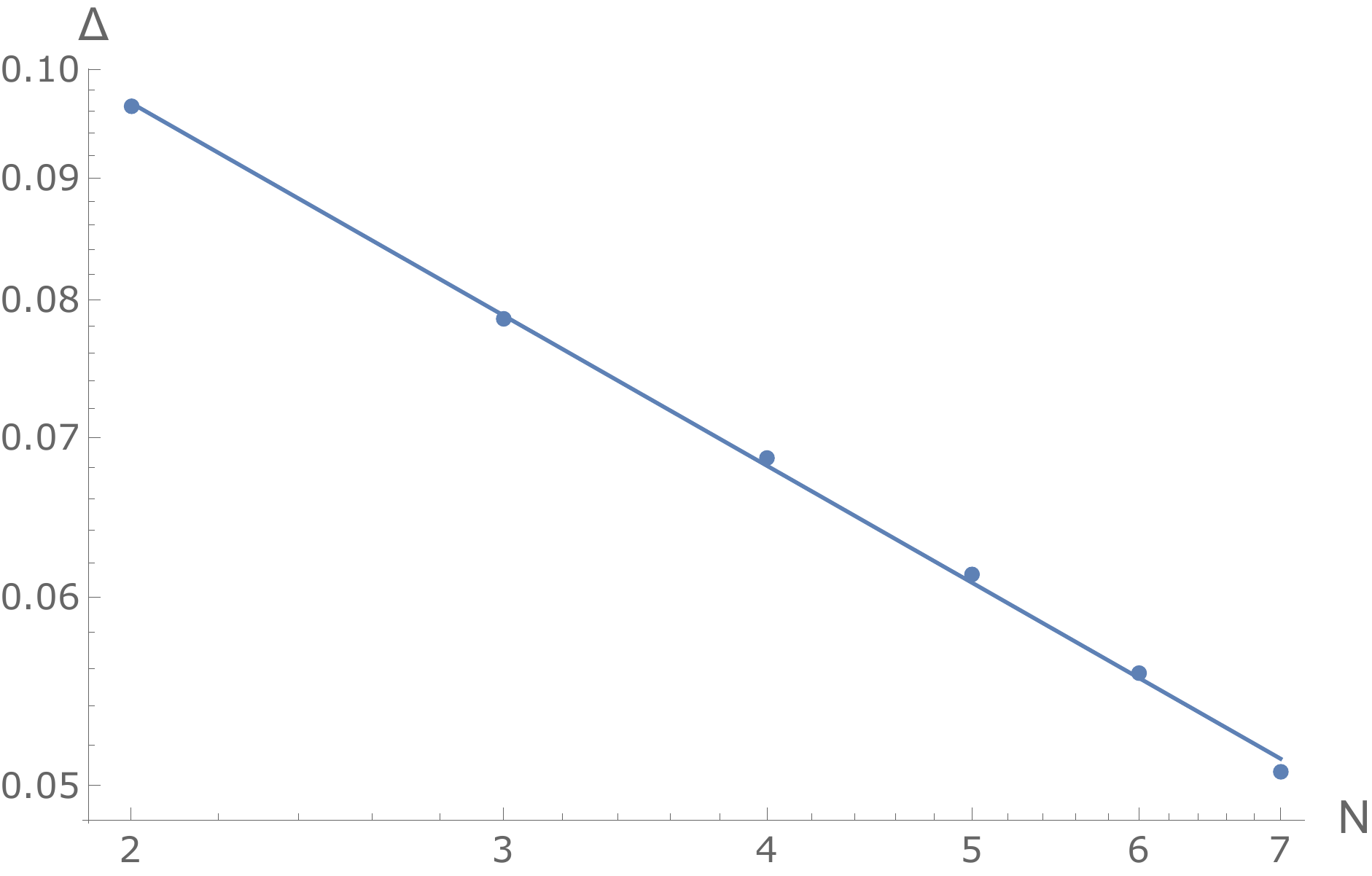}
    \includegraphics[width=3.2in]{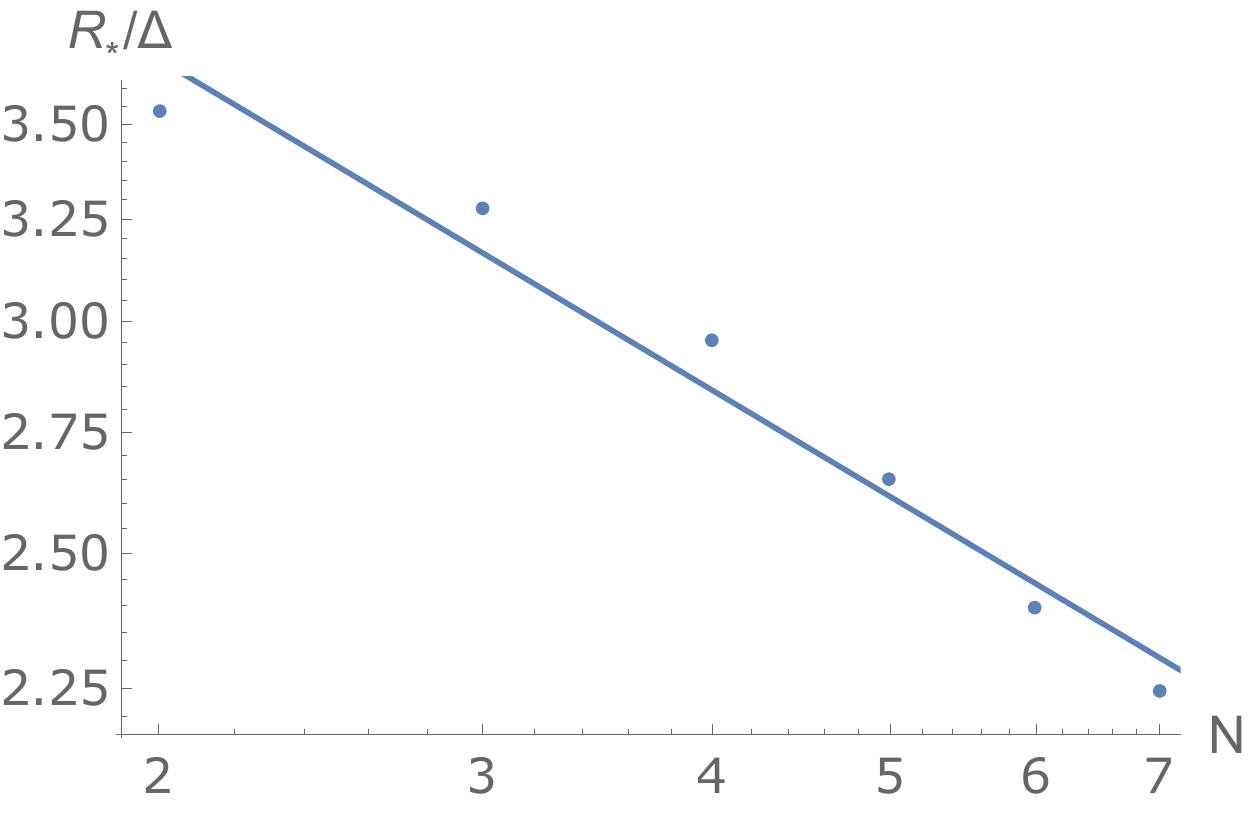}
    \includegraphics[width=3.2in]{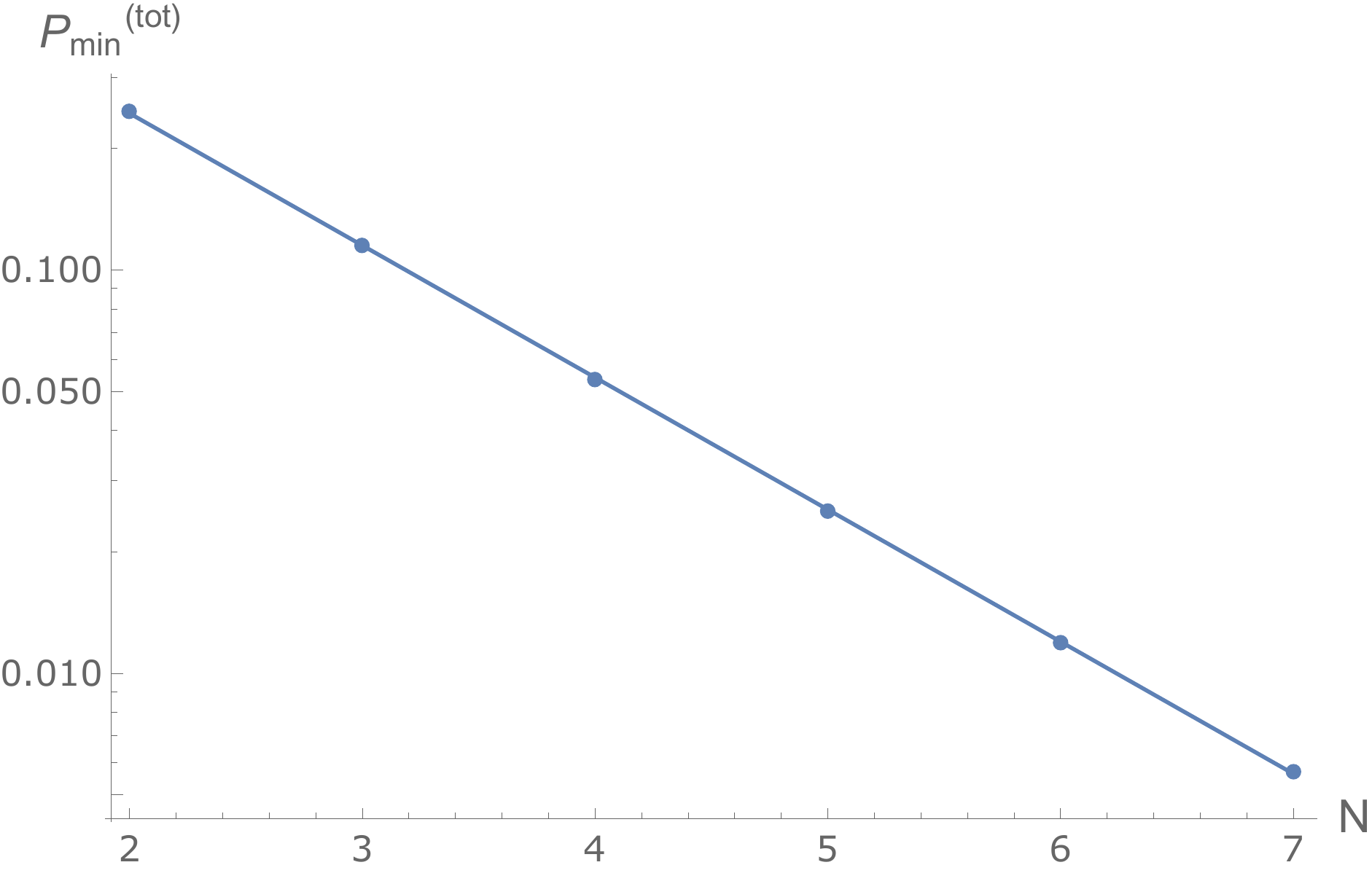}
    \caption{The distributions $P_{\rm min}(R)$ and $P_{\rm max}(R)$ fitted by a hyperbolic tangent (\ref{tanh}).  The blue, orange, green, red, purple and brown  correspond to $N=2, 3, 4, 5, 6$ and 7. The values of $\Delta$ and $R_*/\Delta$ are shown in the middle panel for different number of fields.  The bottom panel is the ratio of the total number of minima to the total number of stationary points, $P_{\rm min}^{(tot)}$.}
   \label{fig:RationVsR}
\end{figure}

The numerically calculated distributions $P_{\rm min}(R)$ are shown in Fig.\ref{fig:RationVsR} for different values of $N$.  We see that as $N$ is increased, the minima are more and more concentrated near $R=0$.  Moreover, for small values of $R$ we have $P_{\rm min}(R)\approx 1$, so almost all stationary points are minima.  Similarly, the maxima of the potential tend to be concentrated near $R=1$, with almost all stationary points being maxima at small values of $(1-R)$.  This is similar to the results found in \cite{BrayDean,Bachlechner} for random Gaussian fields in the large $N$ limit.

The distributions in Fig. \ref{fig:RationVsR} are well fitted by
\beq
P_{\rm min}(R)= \frac{1}{2} \left[ 1+\tanh \left(\frac{R_* -R}{\Delta}\right)\right]
\label{tanh}
\eeq
with $\Delta\propto N^{-0.5}$ and the ratio $R_*/\Delta$ slowly decreasing with $N$.  If this trend were to continue, $P_{\rm min}(R=0)$ would significantly differ from 1 at large $N$.  
However, intuitively one expects the probability of a minimum to approach 1 as $R\to 0$.  Hence, we expect that either the fit (\ref{tanh}) or the behavior of $R_*/\Delta$ should be modified in the large $N$ limit.

The probability for a randomly selected stationary point (at any value of $R$) to be a minimum is given by
\beq
P_{\rm min}^{(tot)}=\frac{{\cal N}_{min}}{{\cal N}_{st}},
\eeq
where ${\cal N}_{min}$ and ${\cal N}_{st}$ are respectively the total numbers of minima and of stationary points.  
Our numerical results for $P_{\rm min}^{(tot)}$ for several values of $N$ are plotted in the bottom panel of Fig.\ref{fig:RationVsR}.  Somewhat surprisingly, they are well approximated by the "naive" formula 
\beq
P_{\rm min}^{(tot)}\approx 2^{-N}.
\label{naive}
\eeq

\begin{figure}[htbp] %  figure placement: here, top, bottom, or page
   \centering
  \includegraphics[width=3.1in]{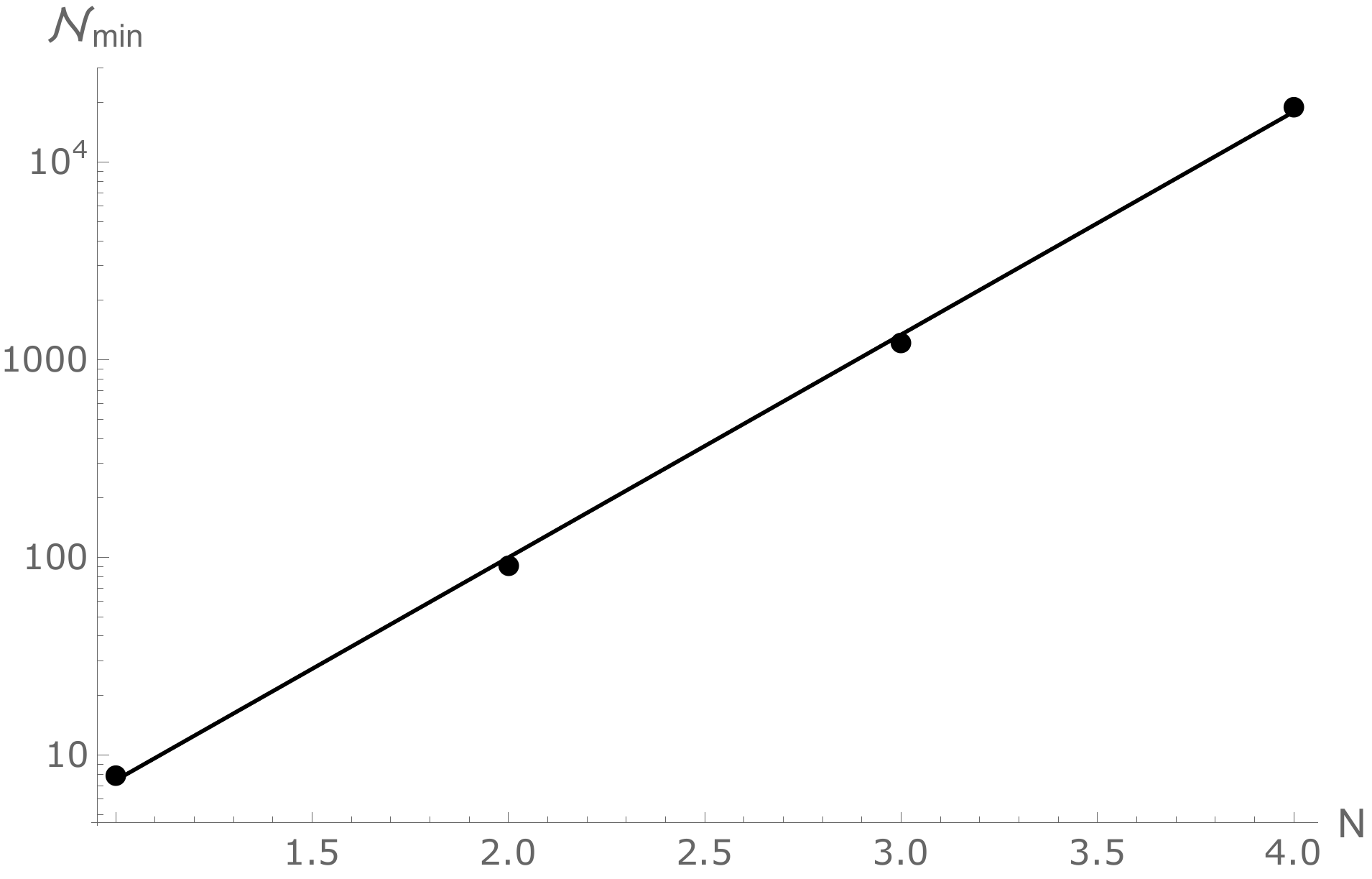} 
  \includegraphics[width=3.3in]{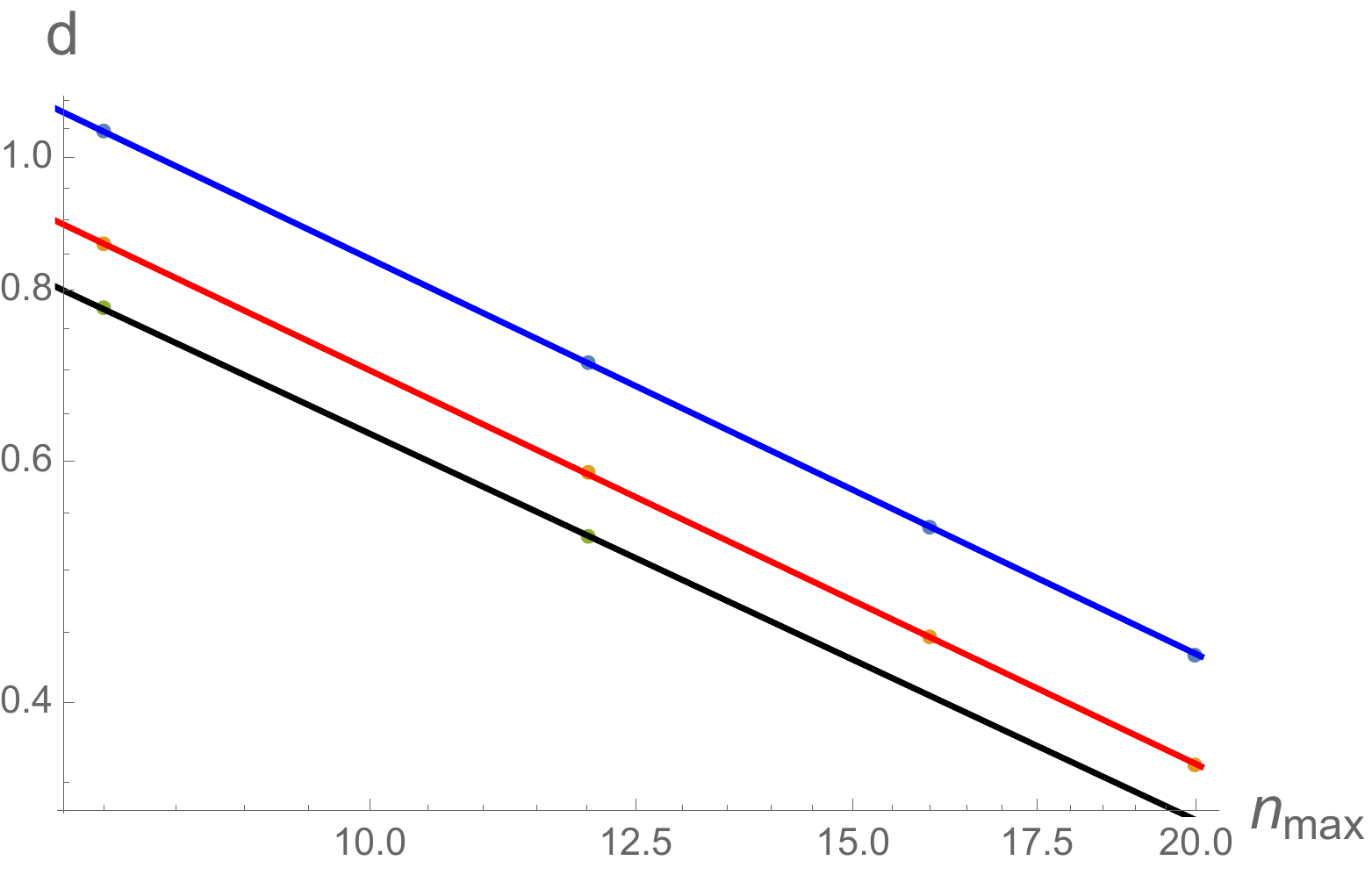} 
   \caption{Left, the average number of minima per realization for $N=1,2,3$ and 4. Right, the typical distance between minima as a function of $n_{\rm max}$. Blue, red and black lines correspond to one, two and three fields.} 
   %The distances were not very sensitive to $N_c$, so we averaged over several values around $N_c=10$. 
   \label{fig:distances}
\end{figure}

The typical distance between the minima, defined as
\beq
d=2\pi {\cal N}_{\rm min}^{-1/N},
\eeq
is plotted in Fig. \ref{fig:distances} as a function of $n_{max}$ for $N=1,2,3$.  As one might expect, it scales as $d\propto n_{max}^{-1}$.  For a fixed value of $n_{max}$, the distance $d$ is a decreasing function of $N$, indicating that the total number of minima grows with $N$ -- even though the probability for a given stationary point to be a minimum rapidly declines.\footnote{This is qualitatively the same behavior as was found in \cite{Bachlechner} for random landscapes with potentials bounded from above and below.}  Indeed, the plot of ${\cal  N}_{min}$ vs. $N$ in Fig. \ref{fig:distances} is well fitted by 
\beq 
{\cal N}_{\rm min} \approx e^{\gamma N}
\label{Nmin2}
\eeq
with $\gamma \approx 2.6$.

Eq.~(\ref{Nmin2}) can be understood as follows.  The total number of stationary points is
${\cal N}_{st} \sim n_{max}^{2N}$, and it follows from Eq.~(\ref{naive}) that ${\cal N}_{\rm min}$ can be written in the form (\ref{Nmin2}) with
\beq
\gamma \approx \ln (n_{max}) .
\label{gammanmax}
\eeq
For $n_{max}=10$ this gives $\gamma \approx 2.3$, which is within 10\% of our numerical estimate.

If the offset parameter in the potential (\ref{landscape}) is set to $V_0 =0$, then $V_{min} \approx V_{max}$ in Eq.~(\ref{RDef}), and minima with $R>0.5$ typically correspond to de Sitter vacua.  We find that the number of such minima decreases with $N$ much faster than exponentially (at fixed values of $n_{max}$ and $N_c$).  The Gaussian fit (\ref{Gaussian}) suggests that $P_{\rm min}(R>0.5)$ decreases with $N$ like an error function, while our numerical results indicate that the decrease is somewhat slower, but still faster than exponential.

\section{Vacuum stability}

The vacuum decay rate can generally be expressed as \cite{Coleman,CallanColeman,CdL}
\beq
\Gamma = Ae^{-2B},
\label{Gamma}
\eeq
where $B$ is the tunneling bounce action and the prefactor $A$ is given by a functional determinant of perturbations about the bounce solution.  The magnitude of $\Gamma$ is mostly determined by the action $B$, so we shall limit ourselves to the calculation of $B$ in our numerical analysis.  Furthermore, we shall assume that gravitational effects on vacuum decay can be neglected.  This is generally the case when bubbles nucleate with a radius much smaller than the Hubble radius of the parent vacuum, $r_b \ll H^{-1}$.  In a landscape with a characteristic energy scale $M$, the typical values are\footnote{The bubble radius can be much larger when the tunneling occurs between nearly degenerate vacua.  We assume that such rare occurrences have little effect on the statistical properties that we are interested in.} $r_b\sim M^{-1}$ and $H^{-1}\sim M_p/M^2$, so one can expect gravitational effects to be unimportant for most of the bubbles, as long as $M\ll M_p$. 

\begin{figure}[htbp] %  figure placement: here, top, bottom, or page
   \centering
    \includegraphics[width=3in]{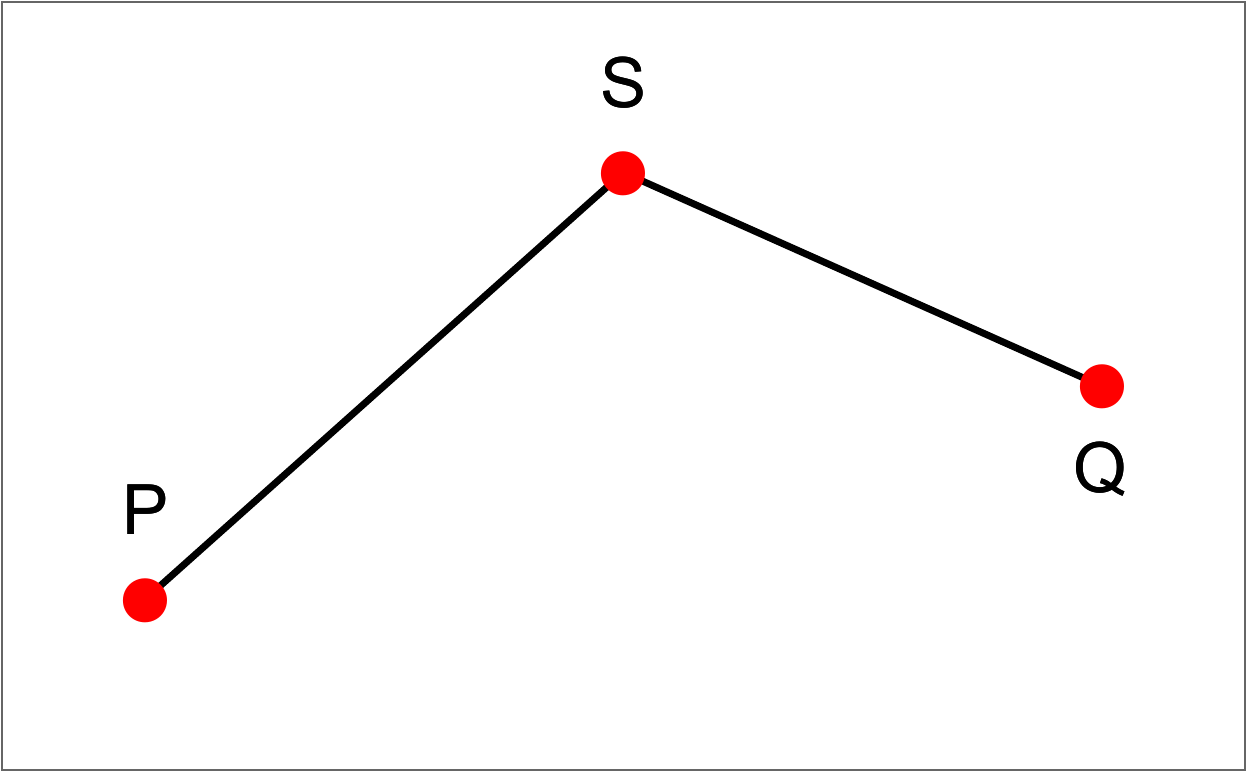} 
    \includegraphics[width=3in]{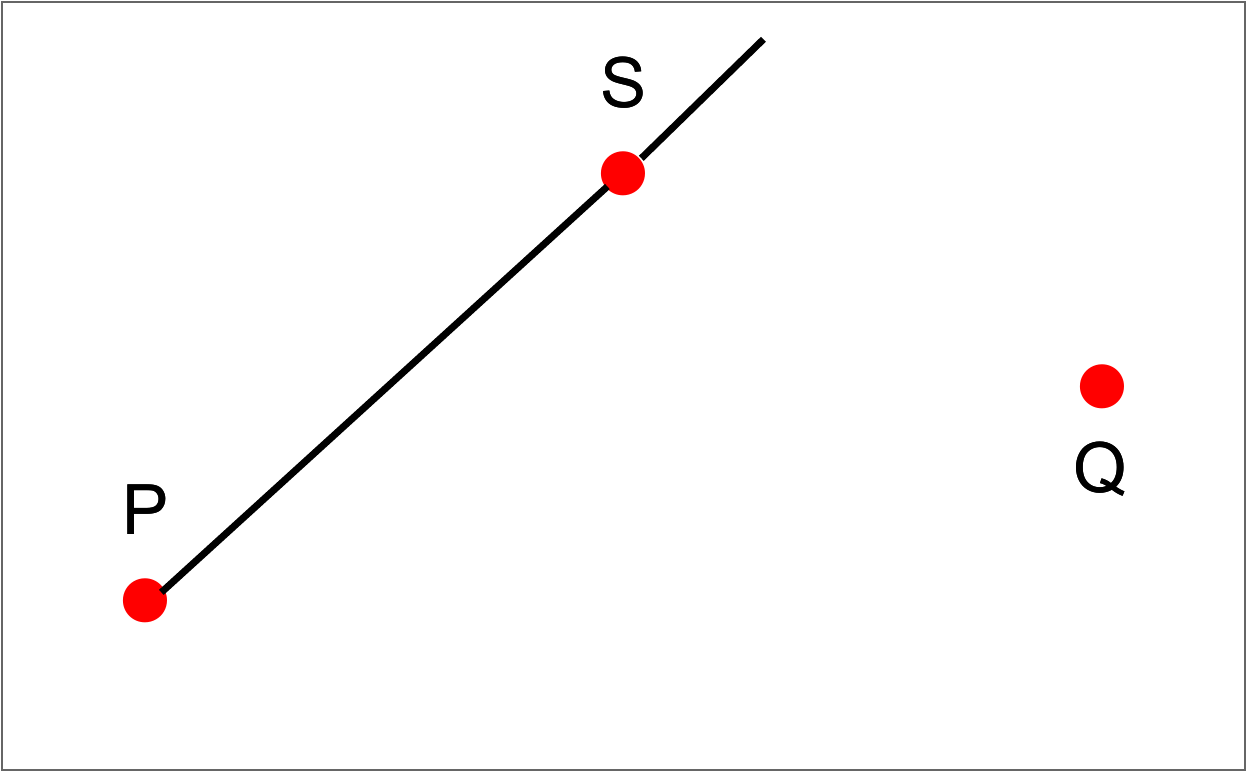}
   \caption{Two possible ways to join vacua $P$ and $Q$. The first, in the left panel, goes straight to the saddle point $S$, then changes the direction and goes straight to the target vacuum. The second path goes straight from $P$ to $S$ and continues in a straight line.}
   \label{fig:twoPaths}
\end{figure}

Finding the bounce solutions of Euclidean field equations is a very complicated numerical problem.  Many instabilities are present and the run-time  grows quickly with the number of fields. Here, we use a proxy to reduce the calculation to a one-dimensional tunneling problem.  
To illustrate the method, suppose we want to find the bounce solution describing tunneling from vacuum $P$ to vacuum $Q$, as shown in Fig.\ref{fig:twoPaths}.  The most probable escape path (MPEP) through the barrier separating the two vacua will typically pass near a saddle point, where the Hessian matrix has a negative eigenvalue in the direction of the path, with all other eigenvalues positive (so the barrier rises as we move away from the MPEP).
Given the two vacua and a suitable saddle point $S$, we approximate the MPEP by two straight segments, the first leading from $P$ to $S$ and the second from $S$ to $Q$, as in the left panel of Fig.\ref{fig:twoPaths}.  There are generally a number of saddle points in the vicinity of vacua $P$ and $Q$; we keep the one which is closest to the line $PQ$ (provided that its projection on $PQ$ is between the points $P$ and $Q$ and that its Hessian matrix has a single negative eigenvalue).  Once we have the approximate escape path $PSQ$, we define the field $\varphi$ along this path as
\beq
d\varphi = \left(\sum_{i=1}^N d\phi_i^2 \right)^{1/2}
\eeq
and find the bounce solution for a one-dimensional tunneling problem in the potential $V(\varphi)$ along this path.  

\begin{figure}[htbp] %  figure placement: here, top, bottom, or page
   \centering
    \includegraphics[width=3in]{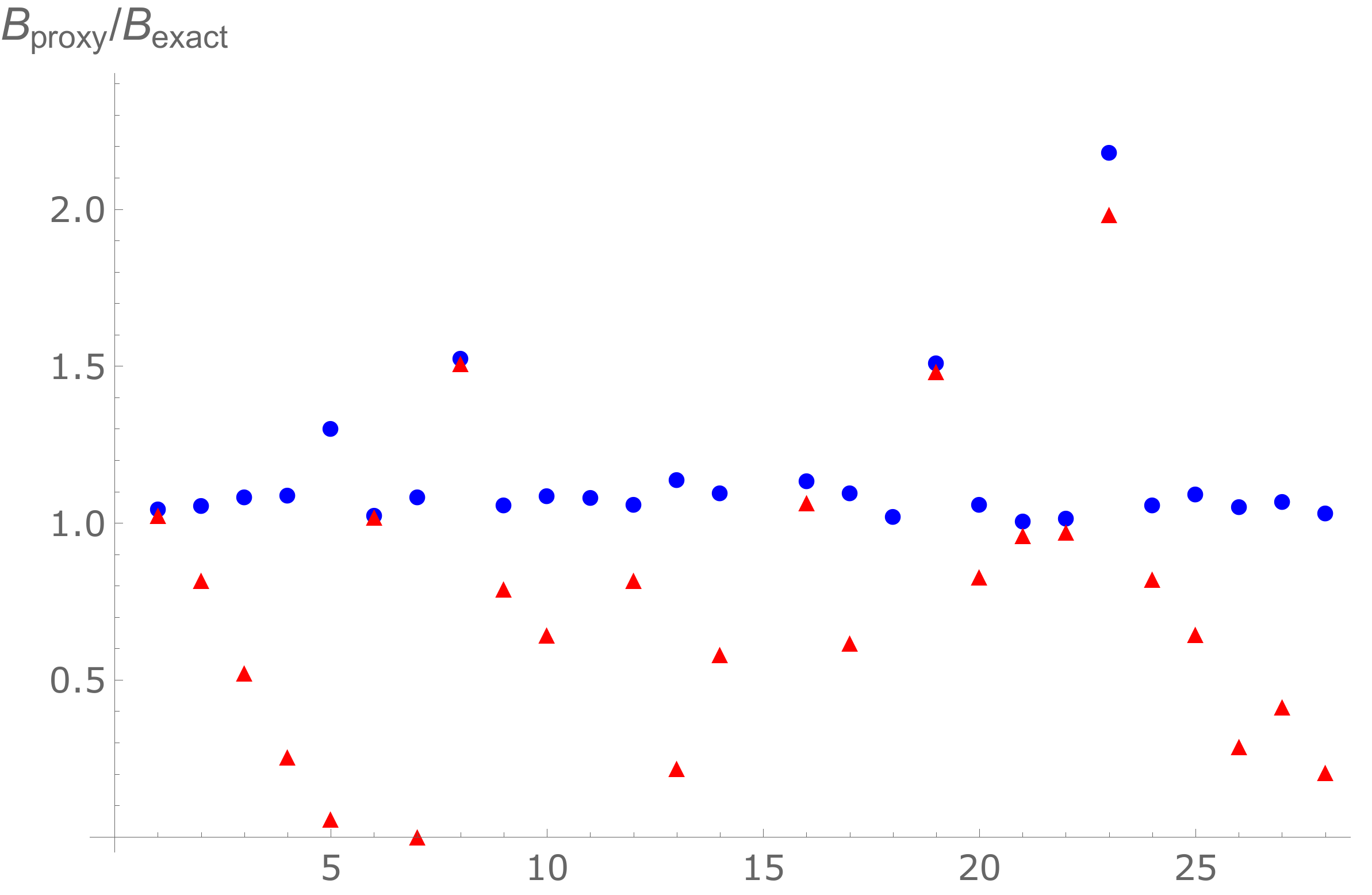} 
   \caption{The ratio of the approximations of the bounce action along path 1 (blue circles) and path 2 (red triangles) to the exact action. We see that the path 1 approximation is very good and we used it as our approximation. }
   \label{fig:compare}
\end{figure}

To test the validity of this $PSQ$ approximation, we compared the bounce actions that it gives with the "exact" actions obtained by numerically solving the field equations using the code developed in Ref. \cite{Ken}.\footnote{In some instances the code of Ref.~\cite{Ken}
fails to find the bounce solution.  We excluded such instances from the sample presented in Fig.\ref{fig:twoPotentials}.  There is a possibility that the performance of the code is correlated with the accuracy of the PSQ approximation; this could introduce bias in our estimate of the accuracy.}  We calculated the bounce actions for tunneling transitions between 27 pairs of vacua in a two-field model with a potential illustrated in Fig.\ref{fig:twoPotentials}.  The ratios of approximate to exact actions for these transitions are shown by blue dots in Fig.\ref{fig:compare}.  Most of these ratios are within 10\% of unity, which is a sufficient accuracy for our statistical analysis.

We also tried an alternative prescription for MPEP: using a straight line from $P$ to $S$ and continuing straight beyond $S$, as shown in the right panel of Fig.\ref{fig:twoPaths}.  The idea is that tunneling takes us from vacuum $P$ through the barrier, and once the field emerges from under the barrier, it evolves classically towards vacuum $Q$.  This suggests that the location of the target vacuum $Q$ may not be important for the bounce solution.  The approximate to exact bounce action ratios obtained using this prescription are marked by red triangles in Fig. \ref{fig:compare}.  The Figure clearly shows that the $PSQ$ prescription is much more accurate than the alternative.  The reason could be that the tunneling endpoint, corresponding to the value of the field at the center of the bubble, tends to be very close to the target vacuum $Q$ \cite{Jun}, indicating that the under-barrier path is indeed close to $PSQ$.

To investigate the statistics of vacuum decay rates, we generated random potentials according to \eqref{landscape} and found all potential minima and all saddle points with one negative Hessian eigenvalue.  For each minimum $P$, we adopted the following procedure for estimating the smallest tunneling action.  First, we find the lower-energy minimum closest to $P$.  Then we identify all neighboring minima (saddle points), which are at most 1.5 times this distance and have  smaller (larger) potential than $V(P)$.  For each neighboring minimum $Q$, we find the saddle point $S$ which is closest to the line $PQ$ and projects on this line somewhere between $P$ and $Q$.  We then calculate the bounce action for all tunneling channels and keep the smallest one.  For global minima or the minima 
having no good saddle points within the specified radius, we assign $B=\infty$ as the tunneling action.  Some further details of our numerical procedure are given in Appendix \ref{numericalBounce}.

\subsection{Numerical results}

\begin{figure}[htbp] %  figure placement: here, top, bottom, or page
   \centering
  \includegraphics[width=3.2in]{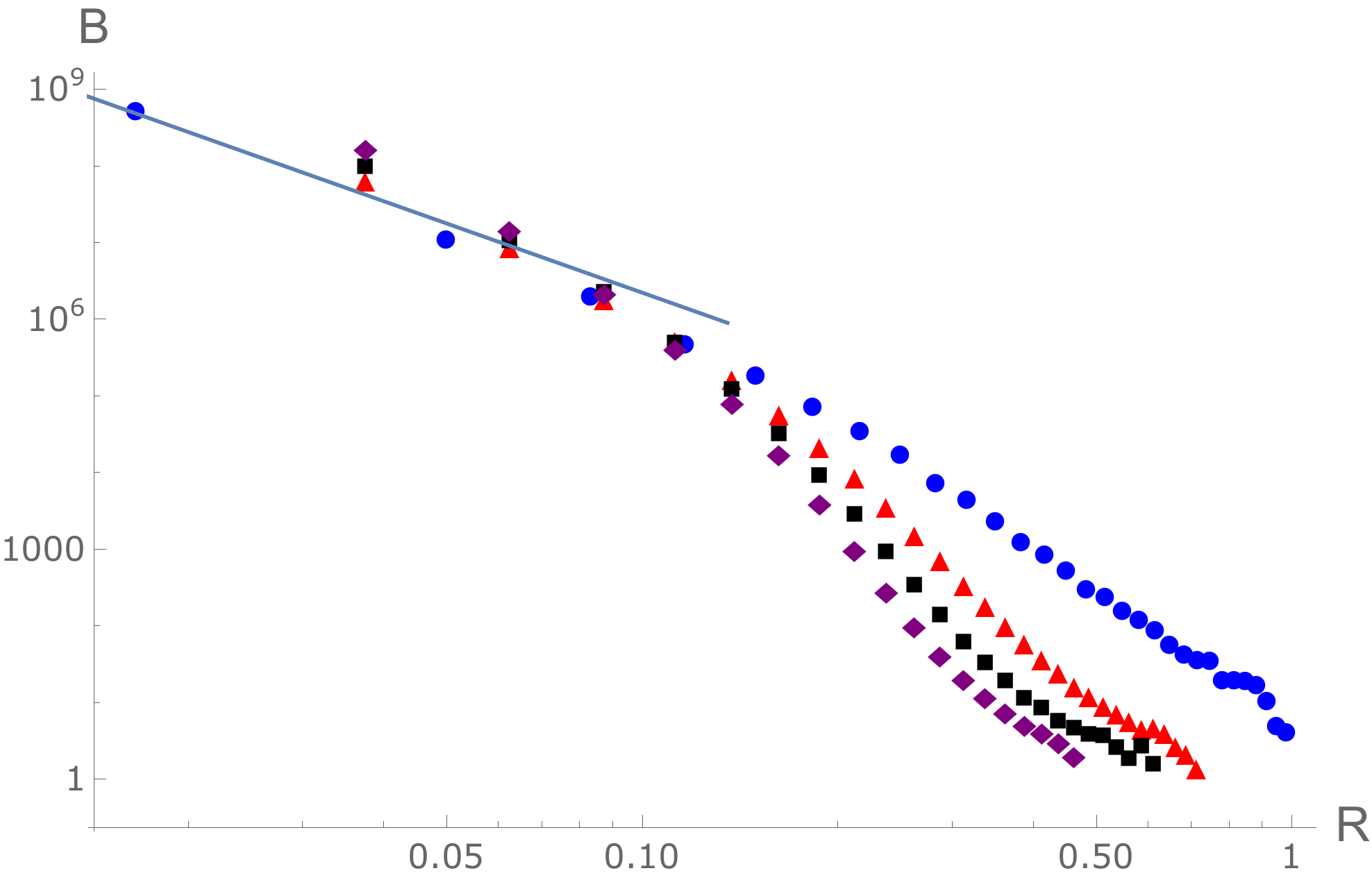} 
%  % \includegraphics[width=3.2in]{1FieldBVsR.pdf} 
%  %  \includegraphics[width=3.2in]{2FieldBVsR.pdf} 
%  %  \includegraphics[width=3.2in]{3FieldBVsR.pdf} 
%  %  \includegraphics[width=3.2in]{4FieldBVsR.pdf} 
\caption{The bounce action $B$ vs. $R$ for $N=1,2,3$ and 4 fields. The thin-wall fit $B\propto R^{-3}$ is also shown as a reference. 
Blue circles, red triangles, black squares and purple diamonds correspond to $N=1,2,3$ and 4. }
   \label{fig:BColemanVsR}
\end{figure}

In Fig.\ref{fig:BColemanVsR}  we plotted the dominant (smallest) tunneling action $B$ as a function of $R$ for $N=1,2,3$ and 4 fields.  More precisely, we split the values of $R$ into small bins and plot the median value of $B$ for each bin.  We see that $B$ is very large at small $R$, but 
decreases rapidly as $R$ is increased.  $B$ also rapidly drops with $N$, except at small values of $R\lesssim 0.1$, where it is roughly independent of $N$.

%For $R\sim 1$ {\bf (We do not need to go to $R\sim 1$ to see that the actions drop rapidly with N. In fact at the place where almost all the minima located, the action has a strong dependence on N)}, $B$ also decreases with $N$, while for $R\lesssim 0.1$ it is roughly independent of $N$ {\bf For larger $N$'s we have to go to much smaller values of $R$ to see the N-independent behavior. Even for N=4, only for $R\sim 0.05$ the  action agrees with other cases. Probably, we can say at very small $R$, the behavior is like $1/R^3$}  

Some of these trends can be understood by considering the expression for the bounce action in the thin wall regime \cite{Coleman},
\beq
B_{thin~wall} \sim 100 \frac{\Sigma^4}{\epsilon^3}.
\label{thinwall}
\eeq
Here, the numerical factor comes from order one numbers and powers of $\pi$, $\epsilon$ is the difference between the energy densities of the two vacua, and $\Sigma$ is the bubble wall tension.  In a landscape with characteristic energy scale $M$ and self-coupling $\lambda$, we have
\beq
\Sigma \sim \Delta V \cdot \delta \sim \lambda^{1/2} M^3
\eeq
and
\beq
\epsilon \sim \Delta V \cdot R ,
\eeq
where $\Delta V \sim \lambda M^4$ is the typical hight of the barrier between the two vacua and $\delta \sim \lambda^{-1/2} M^{-1}$ is the wall thickness.  Substituting all this in (\ref{thinwall}), we have 
\beq
B_{thin~wall} \sim 100 \lambda^{-1} R^{-3}.
\label{thinwall2}
\eeq
The thin wall approximation applies when $\epsilon \ll \Delta V$, that is, when $R\ll 1$.  
The fit in Fig. \ref{fig:BColemanVsR} shows that indeed it gives a reasonable approximation at $R\lesssim 0.1$.

%In the middle of the distribution, that is, for $R$ not too close to 0 or 1, one can expect the vacuum statistics to be qualitatively similar to that in a random Gaussian landscape \cite{??}.  A vacuum can typically decay in $N$ directions in the field space, and the decay rate is dominated by the channel in which the tunneling action $B$ is the smallest.  

%As we see here the tunneling actions drops more or less like a power-law. In Fig.\ref{fig:allBs} we plotted the tunneling exponent for different fields values on the same plot. The tunneling rates drop very quickly as the number of fields increases. 

\begin{figure}[htbp] %  figure placement: here, top, bottom, or page
   \centering
   \includegraphics[width=3.4in]{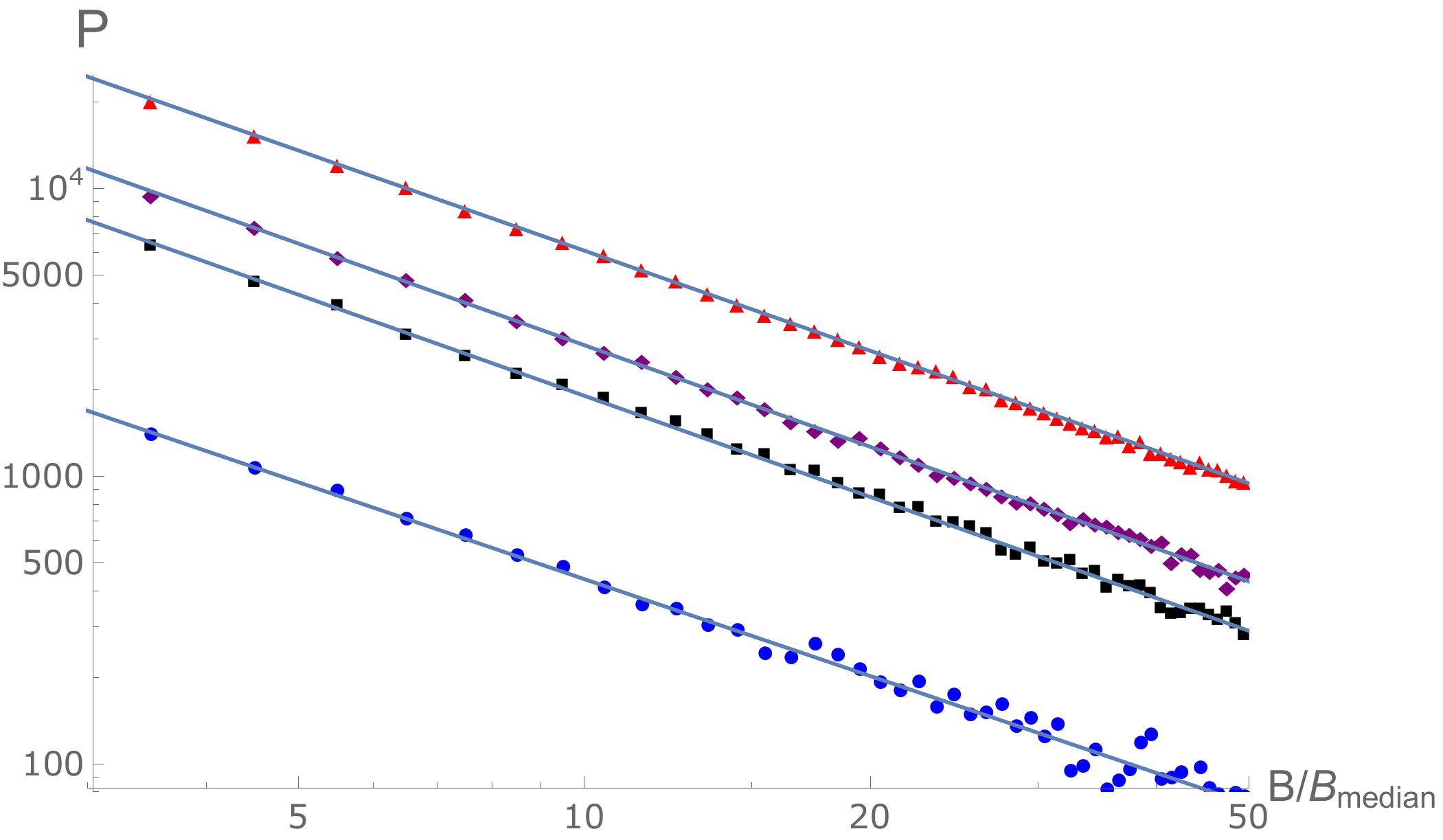} 
   \includegraphics[width=3in]{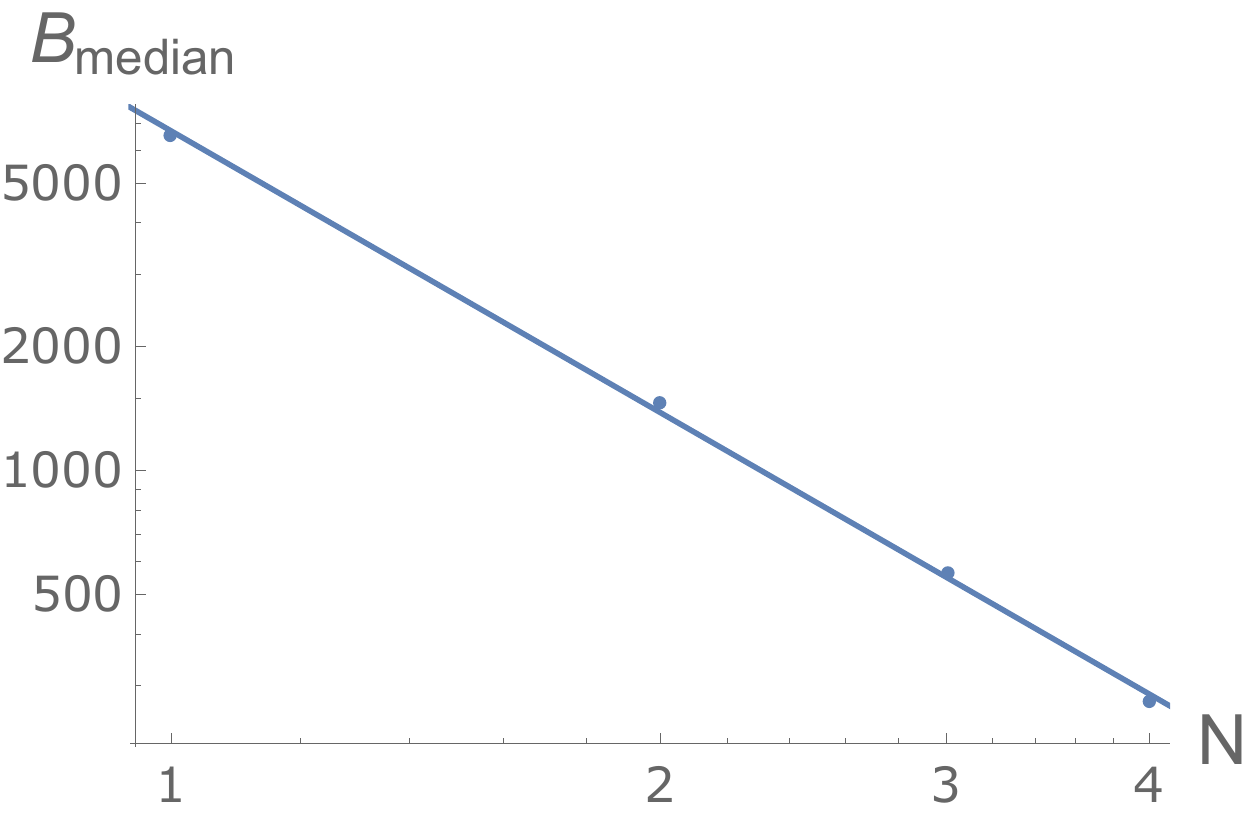}
   \caption{Left, the distribution of the bounce action $B$ at large $B$ for $N=1,2,3$ and 4.  Blue circles, red triangles, black squares and purple diamonds correspond to $N=1, 2, 3$ and 4 fields. 
   %The slopes of all lines are very close to -0.18. 
   The vertical shift of the graphs is due to different sample sizes and does not convey information.  In the right panel, $B_{\rm median}$ is plotted vs. $N$.} 
   %This indicates that $B_{\rm median}\sim N^{-2.3}$.}
   \label{fig:Bdist}
\end{figure}

The median value of $B$ as a function of $N$ is plotted in the right panel of Fig. \ref{fig:Bdist}; it is well fitted by
\bel{medianB}
B_{\rm median} \approx 6000 N^{-2.3} .
\eeq   
The value of $B_{\rm median}$ depends on the overall normalization of the potential.  In our numerical calculations we used the coupling $\lambda=0.1$ in Eq.~(\ref{DeltaA}).  For a general value of $\lambda$, the bounce action is $B\propto \lambda^{-1}$, so we would have
\beq
B_{\rm median} \approx 600 \lambda^{-1} N^{-2.3}.
\label{medianB2}
\eeq   
In realistic landscape models we do not expect $\lambda$ to be particularly small.

The tail of the distribution at $B > B_{\rm median}$, plotted in the left panel of Fig. \ref{fig:Bdist}, is well fitted by 
\bel{tailDist}
%	P(B) \sim e^{-0.18 B/B_{\rm median}}~
	P(B) \sim {\cal A} \left(\frac{B}{B_{\rm median}}\right)^{-q}~
\eeq
with $q\approx 1.15$.  The normalization constant can be estimates as
\beq
{\cal A}\sim \frac{q-1}{2B_{median}}.
\eeq
%Combining \eqref{medianB} and \eqref{tailDist} we get
The fraction of vacua having a bounce action greater than a given value $B\gg B_{median}$ is then
\beq
P_>(B) \sim \frac{1}{2} \left(\frac{B_{median}}{B}\right)^{q-1} .
\label{P>}
\eeq

We note that the power-law distribution (\ref{tailDist}) differs dramatically from the results of  Refs. \cite{Masoumi,Paban}, which found an exponential dependence on $B$ for tunneling in a random quartic potential.  The reasons for this difference are not clear to us; we hope to return to this issue in future work.

We were able to analyze vacuum stability only for relatively small values of $N\leq 4$, so we cannot reach any reliable conclusions for the most interesting case of a large landscape with $N\gg 1$.  The best we can do is to assume that the trends we observe at $N \leq 4$ will continue at larger values of $N$.   Eq.~(\ref{medianB2}) then suggests that for $N\sim 100$ we would have $B_{\rm median} \sim 10^{-2} \lambda^{-1}$.   For a vacuum to survive over the present cosmological timescale $\tau\sim 10^{17}s$, we need $B\gtrsim 100$.  Unless $\lambda$ is very small, most vacua will fall short of this mark, so we can hope to find sufficiently stable vacua only at the tail of the distribution, where $B\gg B_{\rm median}$.
In this regime we can use the distribution (\ref{tailDist}).  The number of long-lived vacua with $B$ greater than a given value can then be estimated as
\beq
{\cal N}_>(B) \sim {\cal N}_{\rm min} P_>(B) \sim N^{-0.3} n_{max}^N (\lambda B)^{-0.14}. 
\label{N>B}
\eeq
where we have used Eqs.~(\ref{Nmin2}) and (\ref{gammanmax}) for the total number of minima.  ${\cal N}_>(B)$ grows rapidly with $n_{max}$ and $N$; we also note the very weak dependence on $\lambda$ and on $B$.  With $n_{max}\gtrsim 10$ and $N\gtrsim 100$ this number may exceed the value $\sim 10^{120}$ necessary to explain the smallness of the cosmological constant.

\section{Models with a mass term}

Our results so far contain some good news and some bad news for the cosine landscape model.  On the positive side, the number of vacua in the model grows exponentially with the number of fields $N$, and even the number of  relatively stable vacua may be sufficient for a successful landscape scenario.  On the other hand, the vacua are concentrated near the absolute minimum $R=0$ in the large $N$ limit.  In fact, the total number of de Sitter vacua (which correspond to $R>0.5$ in models with $V_0=0$) {\it decreases} with $N$, indicating that this number approaches zero as $N\to\infty$.  

We now introduce a modification of the cosine landscape model, where this problem can be addressed.  It has been pointed out in \cite{Gia,Gia2} that an axion-type field $\phi$ with a periodic potential $V(\phi)$ can acquire a quadratic mass term if it interacts with an antisymmetric 4-form field $F_{\mu\nu\sigma\tau}$ via a mixing term
\beq
{\cal L}_{int} = \frac{\mu M}{24} \phi \frac{\epsilon^{\mu\nu\sigma\tau}}{\sqrt{-g}} F_{\mu\nu\sigma\tau} .
\eeq
Here, the coupling $\mu$ has the dimension of mass.  The form field has no dynamical degrees of freedom and can be integrated out, leaving a scalar field with a potential $U(\phi)=V(\phi)+\frac{1}{2} \mu^2 M^2(\phi-\phi_0)^2$, where $\phi_0$ is an integration constant.  We shall set $\phi_0=0$ in what follows.
Multiple axions and form fields are predicted in string theory, and interaction mixing terms are also expected to be present \cite{Gia2,Kaloper}.  In a model including a number of axions $\phi_i$ and form fields $F^a_{\mu\nu\sigma\tau}$, the mixing term is
\beq
{\cal L}_{int} = \frac{M}{24} \sum_{i,a} \mu_{ia} \phi_i \frac{\epsilon^{\mu\nu\sigma\tau}}{\sqrt{-g}} F^a_{\mu\nu\sigma\tau},
\eeq
and the resulting axion mass matrix is 
\beq
M^2_{ij}=M^2 \sum_a \mu_{ia} \mu_{ja}.
\label{M2}
\eeq

We shall assume for simplicity that the mass matrix in (\ref{M2}) is proportional to unit matrix,
\beq
M^2_{ij} = \mu^2 M^2 \delta_{ij},
\eeq
so the potential is
\beq
U(\phi) = V(\phi)+\frac{1}{2}\mu^2 M^2\sum_i \phi_i^2,
\label{U}
\eeq
where $V(\phi)$ is given by Eq.~(\ref{landscape}) with $V_0=0$.  Furthermore, we shall assume, as before, that the kinetic term matrix for the fields $\phi_i$ is given by Eq.~(\ref{Kij}) with $f=M$.  Note that since the fields $\phi_j$ are dimensionless in our convention, the mass matrix elements $M_{ij}$ in (\ref{M2}) have dimension of mass squared.

\begin{figure}[htbp] %  figure placement: here, top, bottom, or page
   \centering
   \includegraphics[width=2.8in]{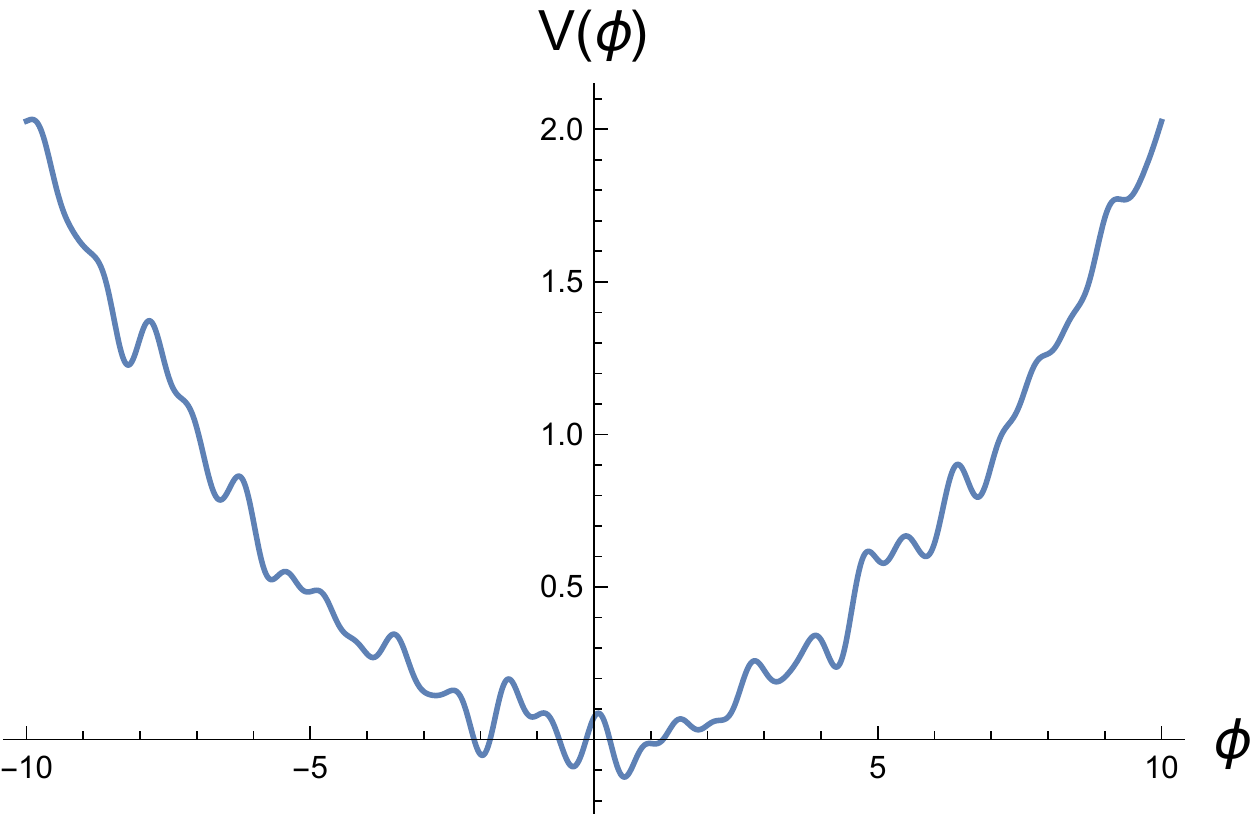} 
    \includegraphics[width=2.8in]{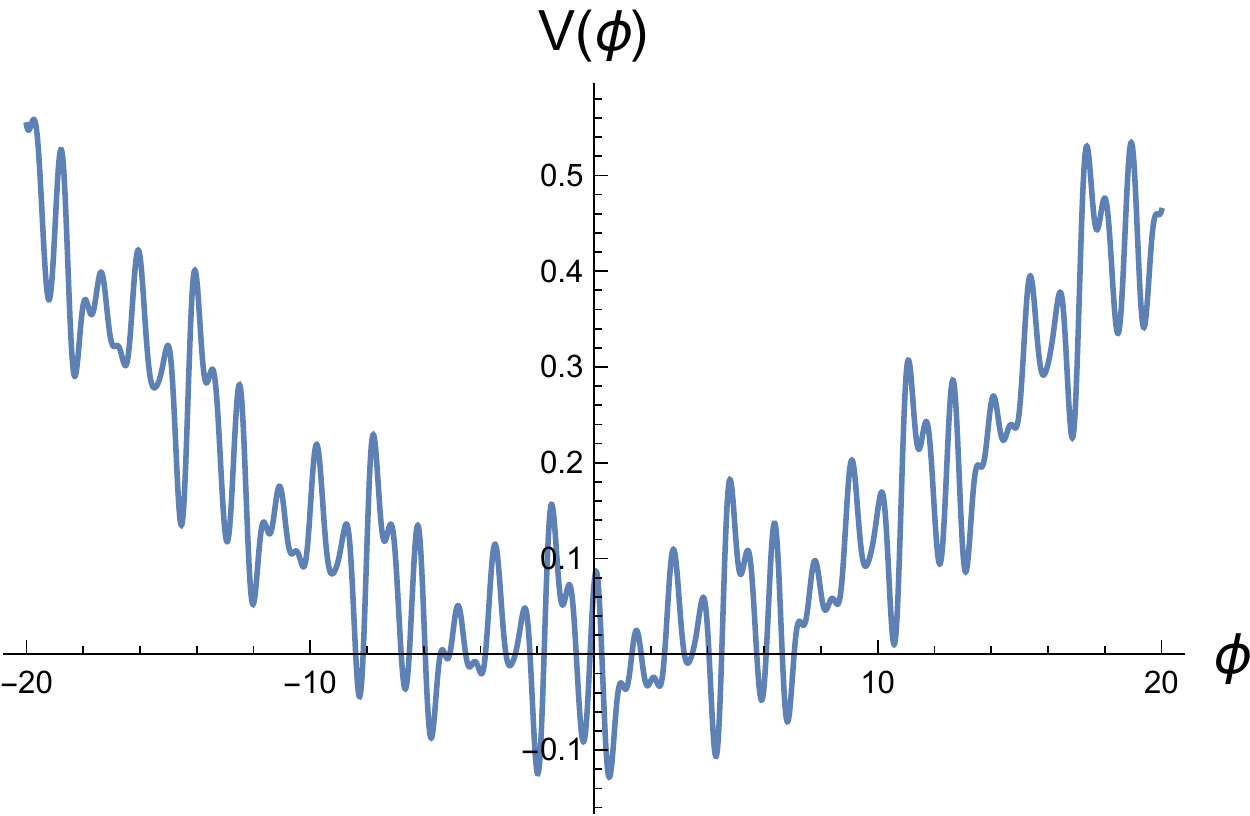} 
   \caption{Examples of potential defined in \eqref{U}. Left panel corresponds to  $\mu=0.2$ and the right panel to $\mu=0.05$.} 
%   Here  we chose $N_c=40$ and $n_{\rm max}=10$. }
   \label{fig:massTermPots}
\end{figure}

The potential (\ref{U}) is no longer periodic.  All of its stationary points are located within a finite range around $\phi_j = 0$,
\beq
|\phi_j| \lesssim \frac{\lambda M^2}{\mu^2} n_{max} \equiv \phi_*.
\label{phi*}
\eeq
Some examples of this potential for $N=1$ are plotted in Fig.\ref{fig:massTermPots}.  The effect of the mass term is to "lift" the vacua at nonzero values of $\phi_i$, so one can expect that the distribution of vacuum energies will not be so concentrated near the bottom.  The typical shift of vacuum energy $U_*$ can be estimated as
\beq
{U_*} \sim \mu^2 M^2 \phi_*^2 \sim n_{max}^2 \frac{\lambda^2 M^6}{\mu^2}.
\label{DeltaU}
\eeq
These estimates assume that the cosine terms in the potential add up with random phases.  Some minima of the potential may occur at $\phi > \phi_*$ when the phases accidentally align, with the full vacuum distribution extending to 
\beq
\phi_{\rm max} \sim \sqrt{N_c}\phi_* , ~~~ U_{\rm max} \sim N_c U_* .
 \label{Umax}
 \eeq

A large number of de Sitter vacua can be expected when $U_*$ is greater than the typical variation of the cosine potential $V(\phi)$, $U_* \gg \lambda M^4$, that is, for
\beq
\mu^2 \ll n_{\rm max}^2 \lambda M^2.
\eeq
With our standard values of $n_{max}=10$ and $\lambda=0.1$, and in units where $M=1$, this gives $\mu\ll 3$.  We shall therefore be mostly interested in small values of $\mu$.

\subsection{Vacuum statistics}

The vacuum distribution at $U\lesssim U_*$ can be estimated as follows.  Consider a potential of the general form 
\beq
U(\phi) = F(\phi) + V(\phi),
\eeq
where $V(\phi)$ is the axionic potential (\ref{landscape}) and $F(\phi)$ is a slowly varying function of $\phi_i$, providing a $\phi$-dependent shift of vacuum energies.  Now consider a thin spherical shell of radius $\phi$ and thickness $d\phi$ in the $N$-dimensional field space.  
We shall assume that $d\phi \gg 2\pi /n_{max}$, so that the shell contains a large number of potential minima.  This number can then be estimated as
\beq
d{\cal N}_{\rm min} \approx \nu_{\rm min} C_{N-1} \phi^{N-1} d\phi ,
\label{shell}
\eeq
where 
\beq
\nu_{\rm min} = \left(\frac{n_{max}}{2\pi}\right)^N
\eeq
is the average density of minima in the potential $V(\phi)$ and 
\beq
C_{N-1} = \frac{2\pi^{N/2}}{\Gamma (N/2)} 
\eeq
is the surface area of a unit $N$-sphere.  Assuming that the minima of $V(\phi)$ are localized within a small range $\Delta V \ll U_*$, the change of vacuum energy across the shell is $dU \approx F'(\phi) d\phi$.  Combining this with Eq.~(\ref{shell}), we find
\beq
\frac{d{\cal N}_{\rm min}}{dU} \approx \nu_{\rm min} C_{N-1} \frac{\phi^{N-1}}{F'(\phi)}.
\eeq
In our case $F(\phi)=\mu^2\phi^2/2$, and thus
\beq
\frac{d{\cal N}_{\rm min}}{dU} \propto \phi^{N-2} \propto U^{\frac{N-2}{2}}.
\label{dNdU}
\eeq

The total number of minima is roughly
\beq
{\cal N}_{\rm min} \sim \frac{C_{N-1}}{N} \left( \frac{n_{max} \phi_*}{2\pi}\right)^N .
\label{calNmin}
\eeq
In the large $N$ limit, with the aid of Stirling formula, this can be approximated as
\beq
{\cal N}_{\rm min} \sim \frac{1}{\sqrt{N}} \left(\frac{e}{2\pi} \frac{n_{max}^2 \phi_*^2}{N} \right)^{N/2},
\eeq
where $e$ is the base of natural logarithm.  Starting with $N\sim 1$, this number grows with $N$, reaches a maximal value
\beq
{\cal N}_{\rm min}^{(max)} \sim e^{\bar N}
\eeq
at $N \sim {\bar N} \equiv n_{max}^2 \phi_*^2 /2\pi$, and then decreases.  For small values of $\mu$, ${\cal N}_{\rm min}^{(max)}$ can be extremely large.

We calculated the vacuum distributions numerically for several values of $\mu$ and $N=1,2,3$.  In preceding sections, we expressed the distributions in terms of the parameter $R$ defined in Eq.~(\ref{RDef}).  Here, it is more convenient to use the parameter
\beq 
{\cal R} = \frac{U-U_{\rm min}}{-2U_{\rm min}}.
\eeq
The reason is that in models with $\mu\neq 0$ the minimum of the potential is about the same as for a massless field, $U_{\rm min}\approx V_{\rm min}<0$, while its maximum is $\mu$-dependent.  In terms of the new parameter ${\cal R}$, the minimum of the potential is at ${\cal R}=0$, and de Sitter vacua correspond to ${\cal R}>0.5$.

\begin{figure}[htbp] %  figure placement: here, top, bottom, or page
   \centering
   \includegraphics[width=3in]{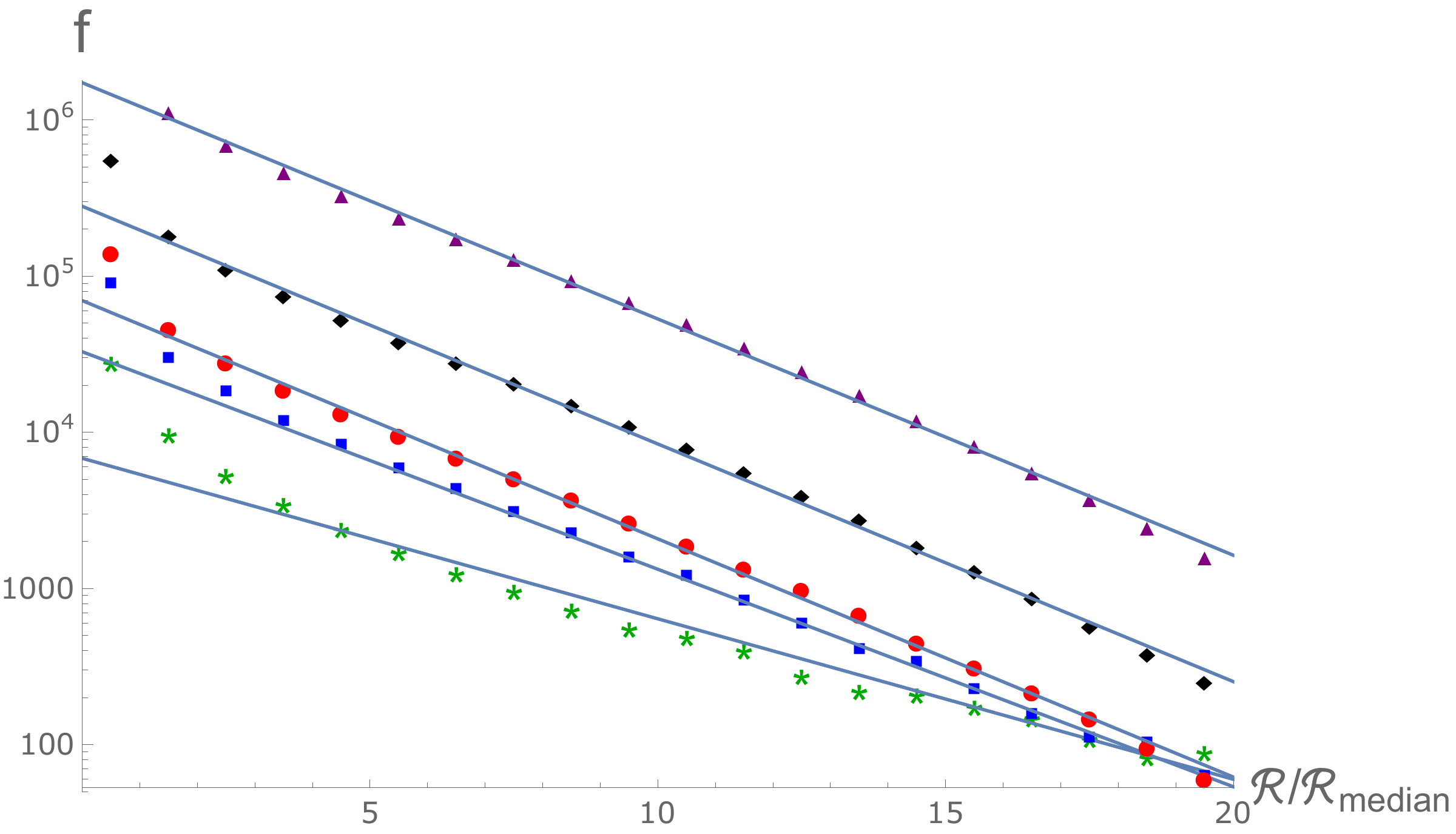}
   \includegraphics[width=3in]{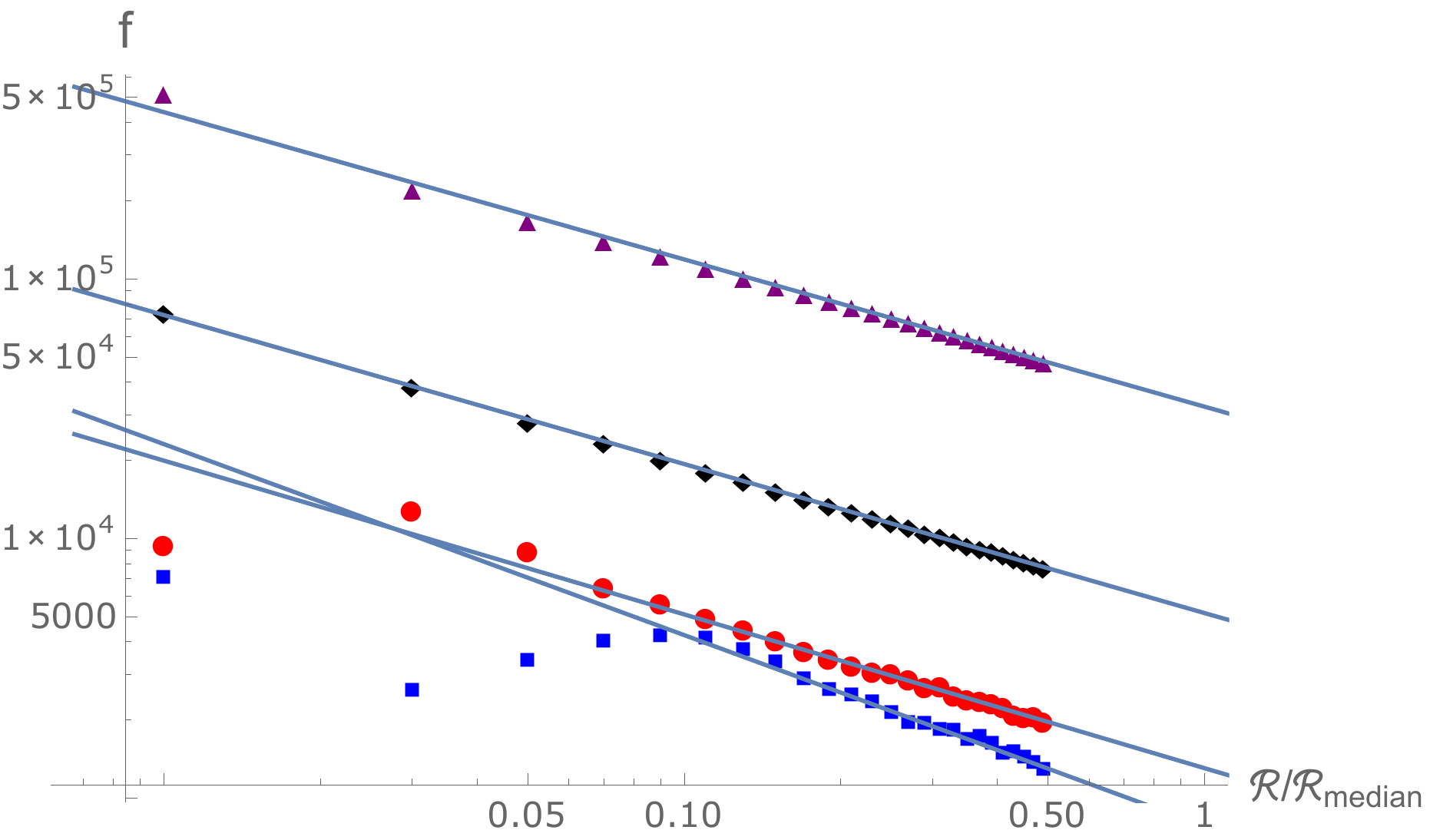}
   \caption{Left, the distribution $f({\cal R/ {\cal R}_{\rm median}})$. 
   %for small values of $\cal R$.  
   %Green, black, red and blue correspond to $\mu=0.05$, 0.02, 0.1 and 0.2. The right panel corresponds to 
   %Right, the tail of the distribution for large values of ${\cal R}$. 
   The blue squares, red dots, black diamonds and purple triangles and green stars correspond respectively to $\mu=0.2$, 0.1, .05, 0.02 and 0.5. We should ignore the vertical shift which is caused by different sample sizes.  The distribution at small values of ${\cal R}$ is shown in the right panel.} 
   %This is consistent with an exponential drop described in \eqref{scriptRFit}.}
   \label{fig:smallRFit}
\end{figure}

\begin{figure}[htbp] %  figure placement: here, top, bottom, or page
   \centering
   \includegraphics[width=3in]{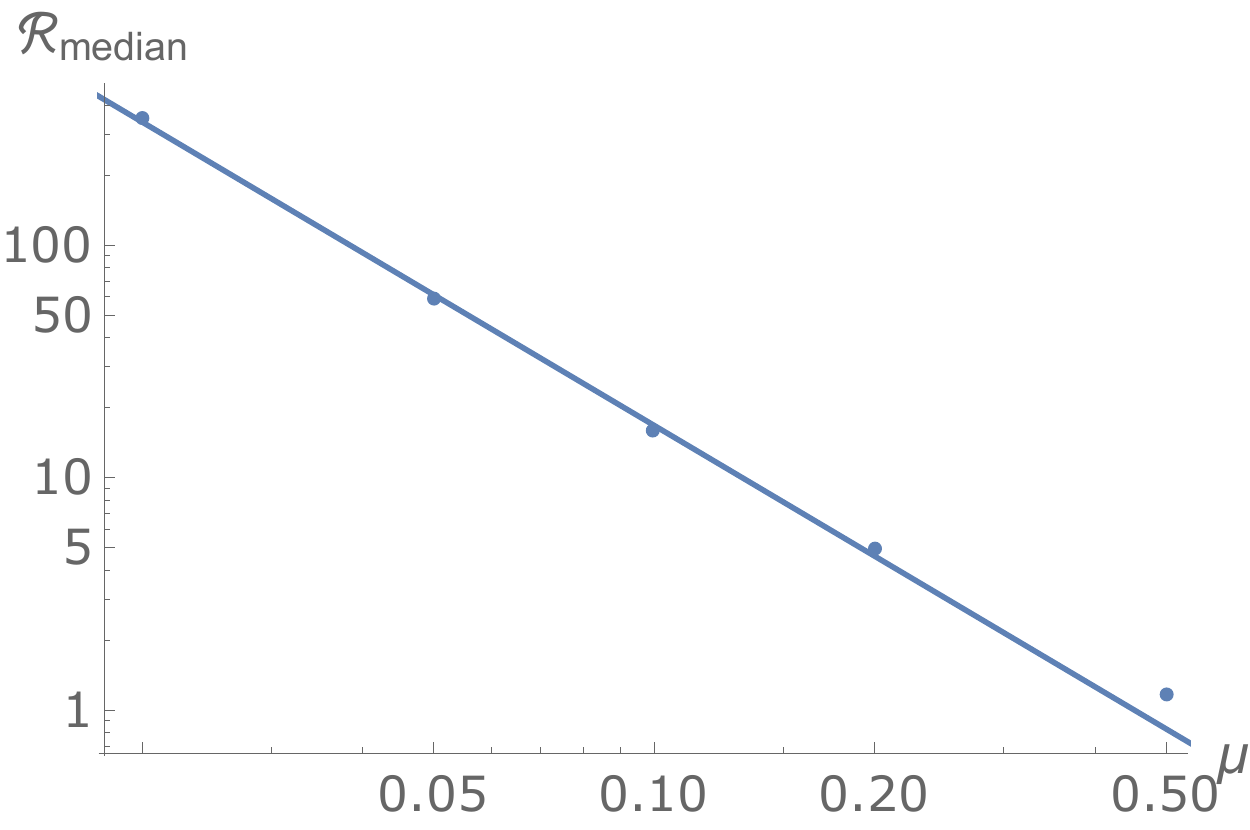} 
   \caption{The median $\cal R$ for different values of $\mu$ for one field.}
   \label{fig:medianFit}
\end{figure}

The numerical vacuum distributions and the median values of ${\cal R}$ for $N=1$ are plotted in Figs.\ref{fig:smallRFit} and \ref{fig:medianFit}, respectively.  The median is fitted by
\beq
{\cal R}_{median} \propto \mu^{-a}
\eeq
with $a\approx 1.9$, which is consistent with the $\mu^{-2}$ dependence of $U_*$ in Eq.~(\ref{DeltaU}).  For all values of $\mu \leq 0.5$ we found that ${\cal R}_{median} > 0.5$, indicating that most of the vacua are de Sitter.

The plots in Fig.~\ref{fig:smallRFit} suggest that the distribution $f({\cal R}/{\cal R}_{median})$ approaches a universal form at small values of $\mu\lesssim 0.05$.  In the small mass regime, the distribution at ${\cal R}\ll {\cal R}_{median}$ is fitted by (see Fig.\ref{fig:smallRFit})
\beq
f({\cal R}) \propto {\cal R}^{-b}
\eeq
with $b = 0.57$, which is reasonably close to the ${\cal R}^{-1/2}$ behavior suggested by Eq.~(\ref{dNdU}) with $N=1$.  For ${\cal R}\gg {\cal R}_{median}$ the distribution is well fitted by an exponential,
\beq\label{scriptRFit}
f({\cal R}) \propto \exp\left( -\kappa {\cal R}/{\cal R}_{median}\right)
\eeq
with the slope $\kappa = 0.35$ for $\mu \lesssim 0.1$ and decreasing towards larger values of $\mu$.  This exponential suppression is due to the fact that values of ${\cal R}\gg {\cal R}_{median}$ require an accidental alignment of phases in a large number of terms of the cosine potential.

\begin{figure}[htbp] %  figure placement: here, top, bottom, or page
   \centering
   \includegraphics[width=4in]{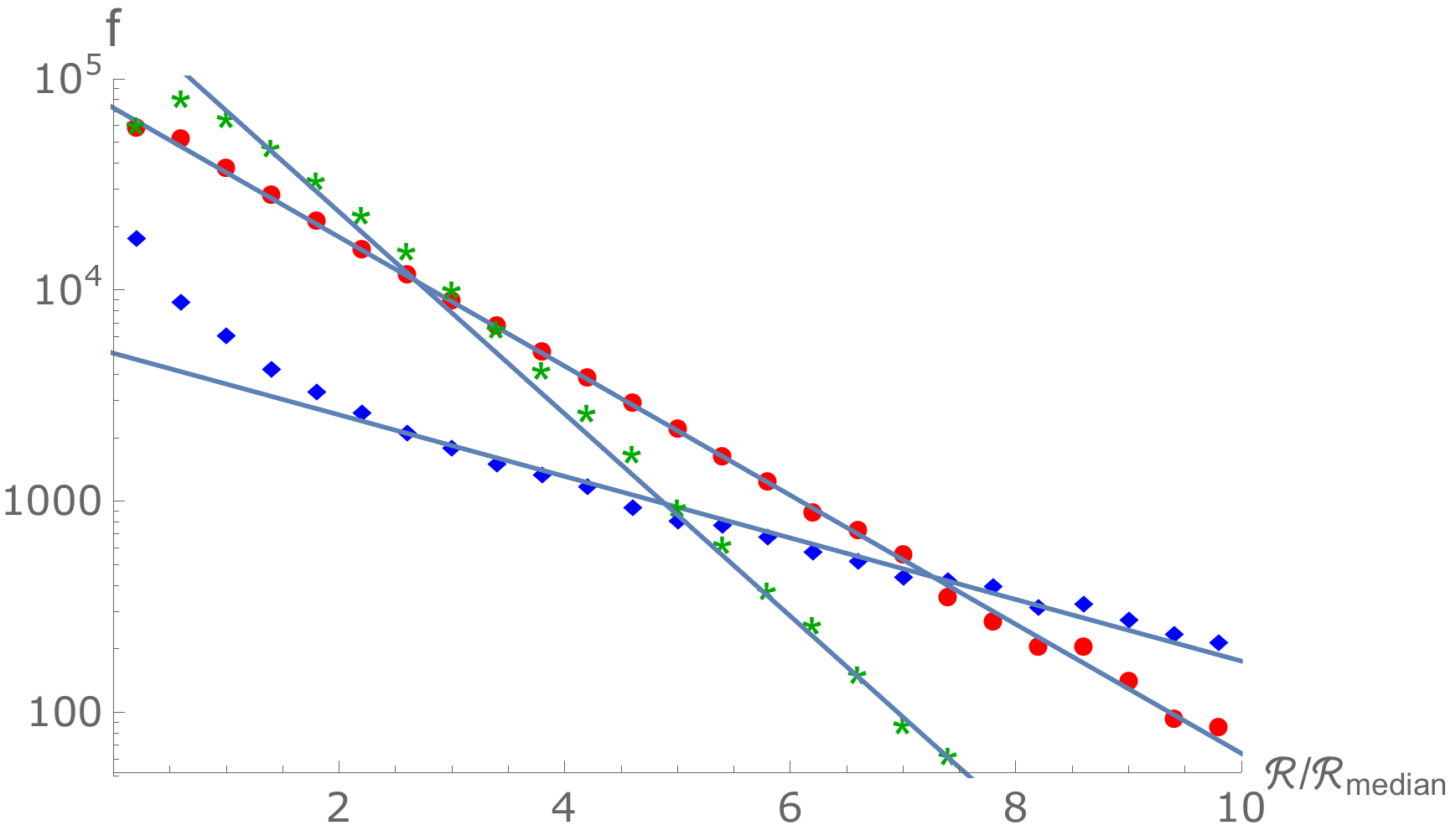} 
   \caption{The distribution of $\cal R$ for $\mu=0.5$ and different number of fields. Blue diamonds, red dots and green stars correspond to $N=1,2$ and 3. At ${\cal R}/{\cal R}_{median} > 2$, the graphs are consistent with the exponential dependence (\ref{scriptRFit}).  
The slope of the three lines are $\kappa=-0.34$, $-0.70$ and $-1.1$.} 
   \label{fig:TailRFitMassive}
\end{figure}

\begin{figure}[htbp] %  figure placement: here, top, bottom, or page
   \centering
   \includegraphics[width=2.3in]{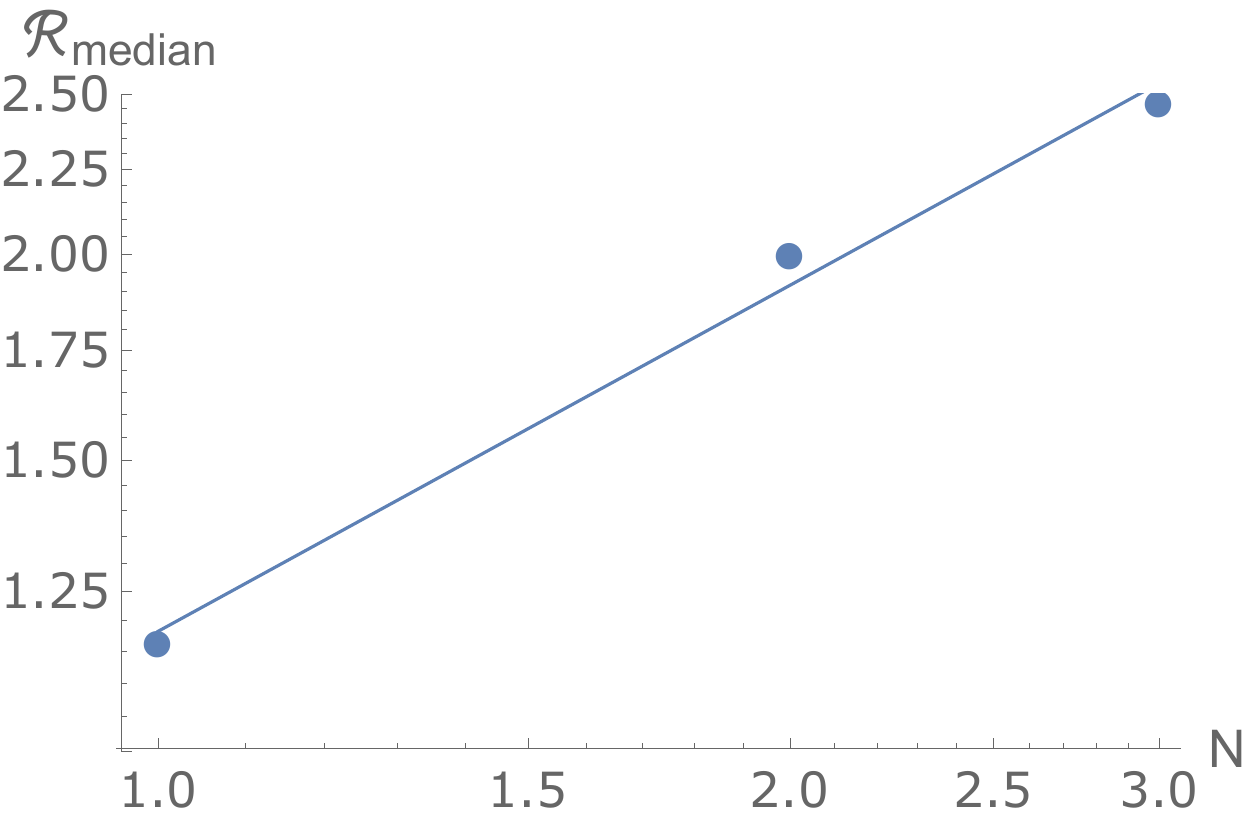} 
   \caption{${\cal R}_{\rm median}$ for $\mu=0.5$ and $N=1,2,3$.} 
   \label{fig:MedianRFitMassive}
\end{figure}

The vacuum distributions for $N=1, 2$ and 3 fields with $\mu=0.5$ are shown in Fig.~\ref{fig:TailRFitMassive}, and the median value of ${\cal R}$ is plotted in Fig. \ref{fig:MedianRFitMassive}.  (Note that the mass $\mu=0.5$ is outside of the range where the analytic expression (\ref{dNdU}) for the vacuum distribution is expected to apply.) 
%but unfortunately the number of vacua at smaller values of $\mu$ is too large for a numerical study.  
We see from Fig. \ref{fig:MedianRFitMassive} that ${\cal R}_{median}$ increases with $N$, suggesting that at large $N$ a large fraction of vacua are de Sitter.    At ${\cal R} > {\cal R}_{median}$, the distribution declines exponentially, as in Eq.~(\ref{scriptRFit}), with the slope getting steeper at larger $N$.  (The plots are consistent with power-law dependences ${\cal R}_{median} \propto N^{0.7}$ and $\kappa = 0.34 N$, but of course these fits based on just three points should not be taken very seriously.)

\subsection{Vacuum stability}

We now consider the effect of a nonzero mass term on vacuum stability.  At $\phi \ll \phi_*$ (or $U\ll U_*$), the gradients of the potential are dominated by the cosine terms, so the mass term has very little effect on the shapes of potential barriers and on the bounce actions.  Hence, in this regime we expect the distributions $P(B)$ and $P_>(B)$ to be roughly independent of the vacuum energy $U$.  The number of vacua with $B$ greater than a given value can then be roughly estimated as
\beq
{\cal N}_>(B) \sim {\cal N}_{\rm min} P_>(B) ,
\eeq
with ${\cal N}_{\rm min}$ from Eq.~(\ref{calNmin}) and $P_>(B)$ from Eqs.~(\ref{P>}), (\ref{medianB2}).  For $B,N\sim 100$ and $n_{min}\sim 10$, this number can be enormous, especially for small values of $\mu$.

\begin{figure}[htbp] %  figure placement: here, top, bottom, or page
   \centering
   \includegraphics[width=3.2in]{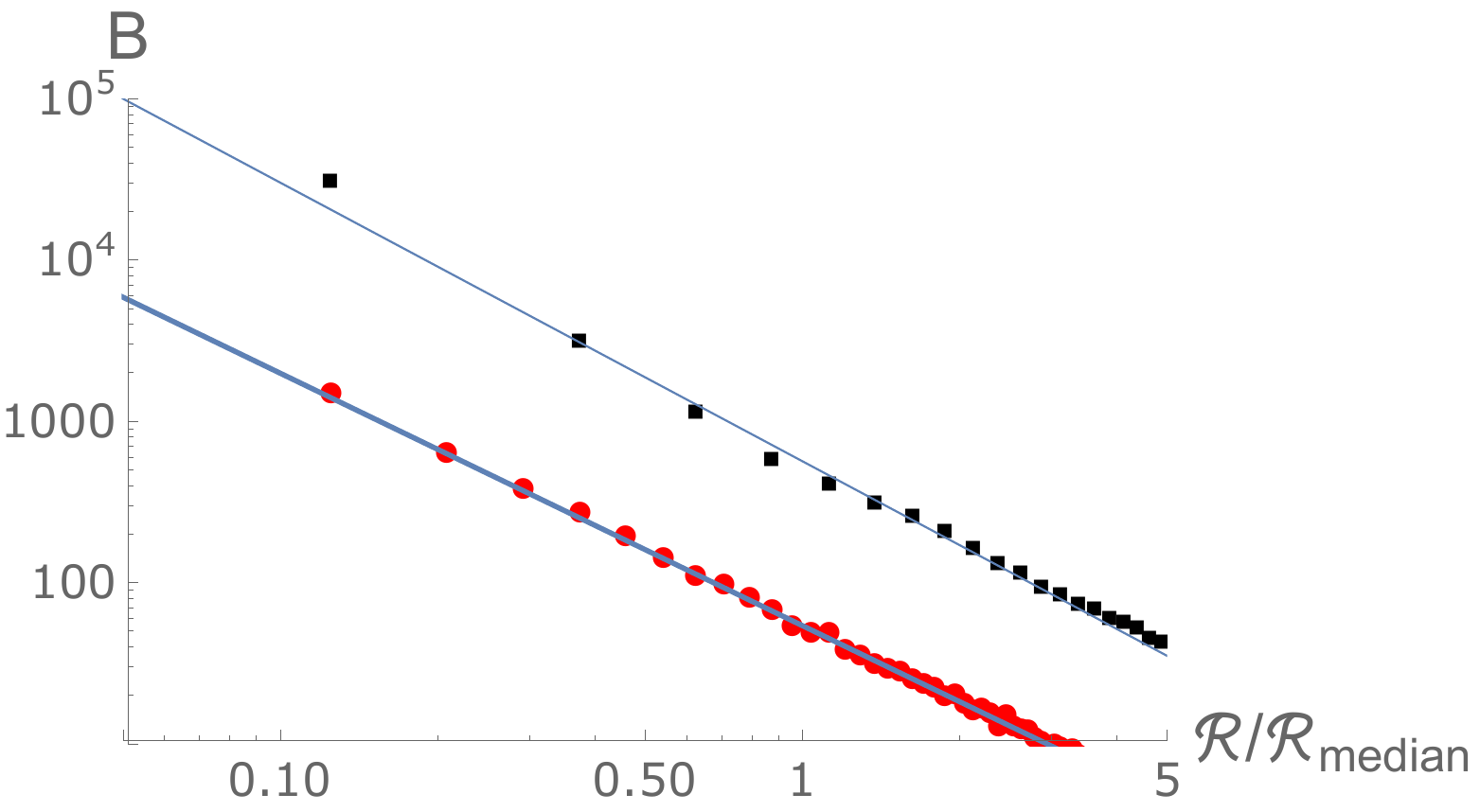} 
   \includegraphics[width=3.2in]{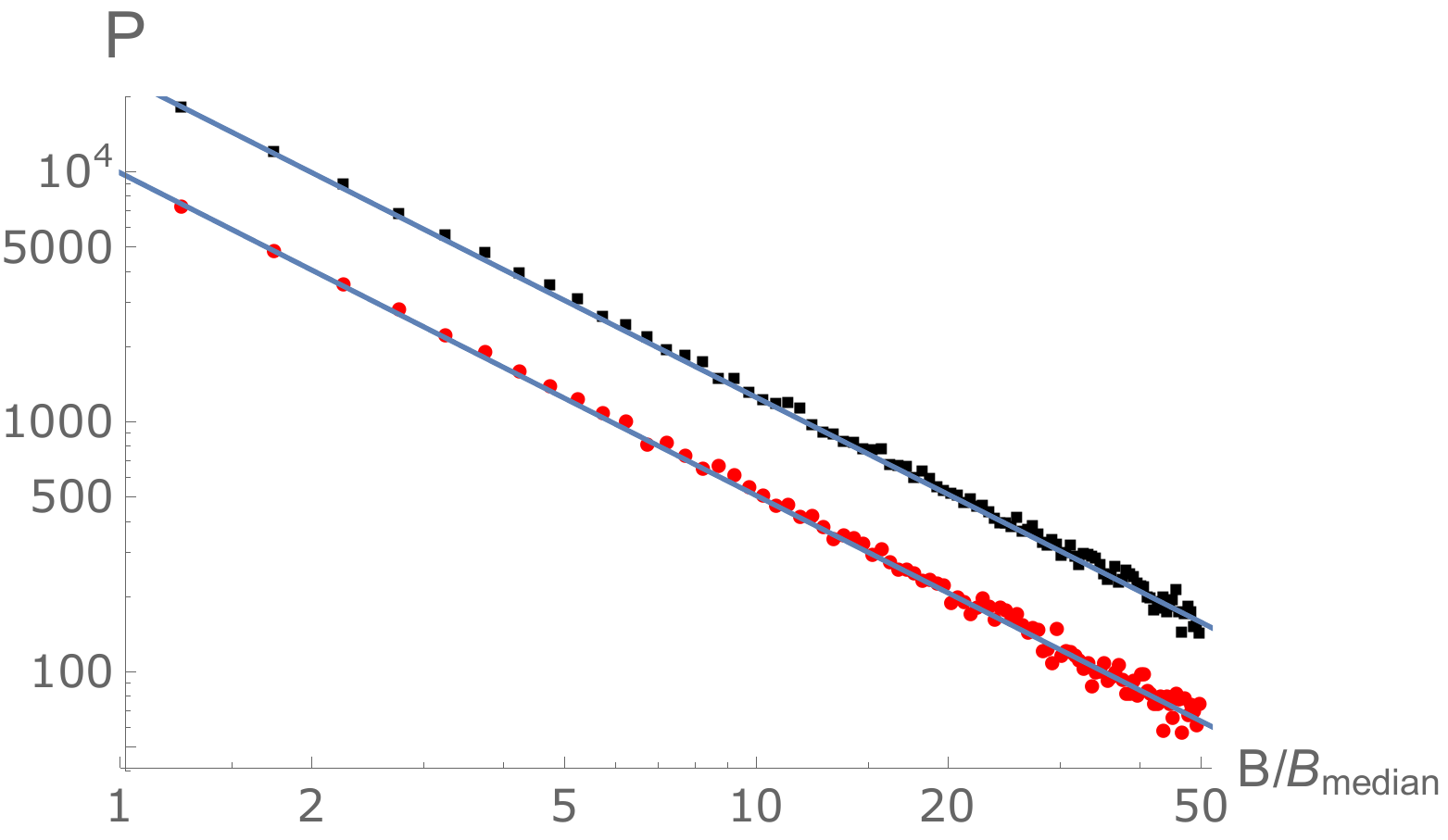} 
    \caption{Left panel is a plot of the bounce action $B$ vs. ${\cal R}$.
      The right panel  shows the distribution $(P(B)$. The black diamond and red dots (on both panels) correspond to $\mu=0.2$ and $0.1$ }
   \label{fig:DecayRateMassive}
\end{figure}

Because of the large run-times involved, we did numerical calculations only for the case of $N=1$ with $\mu = 0.1$ and 0.2.  The tunneling action $B$ is plotted in Fig.\ref{fig:DecayRateMassive} as a function of ${\cal R}$; it is fitted by a power law
\beq
 B \propto \left(\frac{\cal R}{{\cal R}_{median}}\right)^{-\beta}
\eeq
with $\beta \approx 1.6$.  The distribution $P(B)$ is plotted in Fig. \ref{fig:DecayRateMassive}.  It is also well fitted by a power law
\beq
P(B) \propto \left(\frac{B}{{B}_{median}}\right)^{-q}
\eeq
with $q\approx 1.3$.  This is reasonably close to the $\mu=0$ distribution (\ref{tailDist}), as expected.

The median value of $B$ is $B_{median} \approx 40$ for both $\mu=0.1$ and 0.2.  This may look surprisingly small compared to the median for the massless case, Eq.~(\ref{medianB}).
The reason is that the massive models have large numbers of vacua of lower stability at $U\gtrsim U_*$.

\section{Conclusions}

This work was motivated by recent results suggesting serious problems with large vacuum landscape models.  Analysis of vacuum decay in random potentials in Ref.~\cite{Masoumi} suggested that vacuum stability rapidly deteriorates when the number of fields $N$ is increased, so that the number of sufficiently stable vacua in a large landscape is too small to solve the cosmological constant problem and may even be too small to account (without fine-tuning) for the current age of the universe.  Moreover, studies of vacuum statistics in Refs.~\cite{BrayDean,Bachlechner,Battefeld} indicate that with increasing $N$ vacua tend to concentrate more and more near the absolute minimum of the potential, suggesting that nearly all vacua in a large landscape are AdS.

Here, we studied vacuum statistics and stability in axionic landscapes with a potential of the form (\ref{landscape}).  We characterized the stability of a vacuum by the tunneling action $B$ for its decay.  We found that $B$ is strongly correlated with the vacuum energy density, with $B$ rapidly growing (and thus stability increasing) as the energy density is decreased.  The numerically calculated probability for a randomly picked vacuum to have $B$ greater than a given value is well fitted by a power law, $P(B)\propto B^{-0.15}$.  This surprisingly slow decline is in sharp contrast with the analyses of random quartic potentials \cite{Masoumi}, which found an exponential dependence, $P(B) \propto \exp(-KB)$, with $K$ growing as a power of $N$.  

We were able to perform our numerical analysis of vacuum stability only for a relatively small number of fields, $N\leq 4$.  Assuming that the trends we found in this range extend to larger values of $N$, the total number of vacua with $B$ greater than a specified value grows very rapidly with $N$ and is rather insensitive to $B$ (see Eq.~(\ref{N>B})).  With $N\sim 100$, the number of relatively stable vacua may reach the values $\gtrsim 10^{120}$ required for solving the cosmological constant problem.  However, these stable vacua are concentrated near the minimum of the potential, so in order to account for the observed value of the vacuum energy density, one has to fine-tune the offset parameter $V_0$ in (\ref{landscape}) so that the minimum is at $V_{\rm min}\approx 0$.  Thus, this kind of landscape does not provide a solution to the cosmological constant problem.

To address this difficulty, we considered a modification of the model, where the axions acquire a quadratic mass term, due to their mixing with 4-form fields.  This results in a much broader distribution of vacuum energies.  With zero energy offset ($V_0 =0$), for small enough values of the mass parameter we find that most of the vacua in the landscape are de Sitter.  One might be concerned that this kind of models could suffer from the opposite problem of having predominantly de Sitter vacua with large vacuum energies.  This can be addressed by introducing a large negative offset, $V_0 <0$, as it was done in Ref.~\cite{BP}.  The total number of relatively stable vacua in the massive models can be extremely large, especially for small values of the mass parameter.

Apart from the low vacuum energy density and high stability, a successful landscape model should account for a period of slow-roll inflation.  The conditions for inflation in axionic landscapes have been discussed in Refs. \cite{Higaki,Higaki2}.  It would be interesting to study the probability distribution for the number of e-folds of inflation in such models and its possible correlation with the vacuum energy density and stability.  We hope to address this issue in future work.

\section*{Acknowledgements}

This work was supported in part by the National Science Foundation.   We are grateful to 
Thomas Bachlechner, Thorsten Battefeld, Christopher Burke, Eric Roebuck, and especially to Ken Olum for useful discussions and comments.

\appendix

\section{Numerical techniques for finding the critical points and minima} \label{sec:findRoot}
Numerical methods cannot guarantee (except for special case of polynomial equations) finding all the critical points of a function. Equation solving methods may be unable to find minima which have specific features like the ones which are very narrow and deep  and these minima  are likely to be hidden from the solver. This in turn causes bias in the statistics of the stationary points. Fortunately, the function we deal with in \eqref{landscape} has a finite bandwidth in Fourier space (largest $k$ is given by $n_{\rm max})$ and cannot vary very quickly. Therefore, if we start from a dense lattice of initial points  for an equation solver, we can more or less be sure that we find all the stationary points. The lattice spacing should be smaller than a quarter of the smallest wavelength. For example in the above examples where we had 3 fields and $n_{\rm max}=10$, we had to start from order $10^5$ points to make sure we have all the stationary points. This procedure becomes impractical for more than 4 or 5 fields. The number of stationary points, even for small $n_{\rm max}$, grows very rapidly with $N$. 
%For example for 10 fields and $n_{\rm max }=3$ we  need a grid of $10^{12}$ points to assure finding all the minima and stationary points, which is not practical. 
Therefore, we use a Monte-Carlo simulation to find the distribution of the minima for $N>4$. There are two possibilities: 
\begin{enumerate}
	\item We  solve the equations $\vec \nabla V=0$ starting from a grid of random initial points. This way we find all types of stationary points. However, the chance of getting a minimum among them drops as (probably) $1/2^{N}$.
	\item
	We start from a set of random points and find minima close to them. The main advantage is that we do not need to find a large set of stationary points before finding a minimum. The disadvantage is that we lose the statistics of minima vs stationary points. 
\end{enumerate}
Because we were also interested in the statistics of the stationary points we used the first method (which is more time-consuming). 

\begin{figure}[htbp] %  figure placement: here, top, bottom, or page
   \centering
   \includegraphics[width=3.2in]{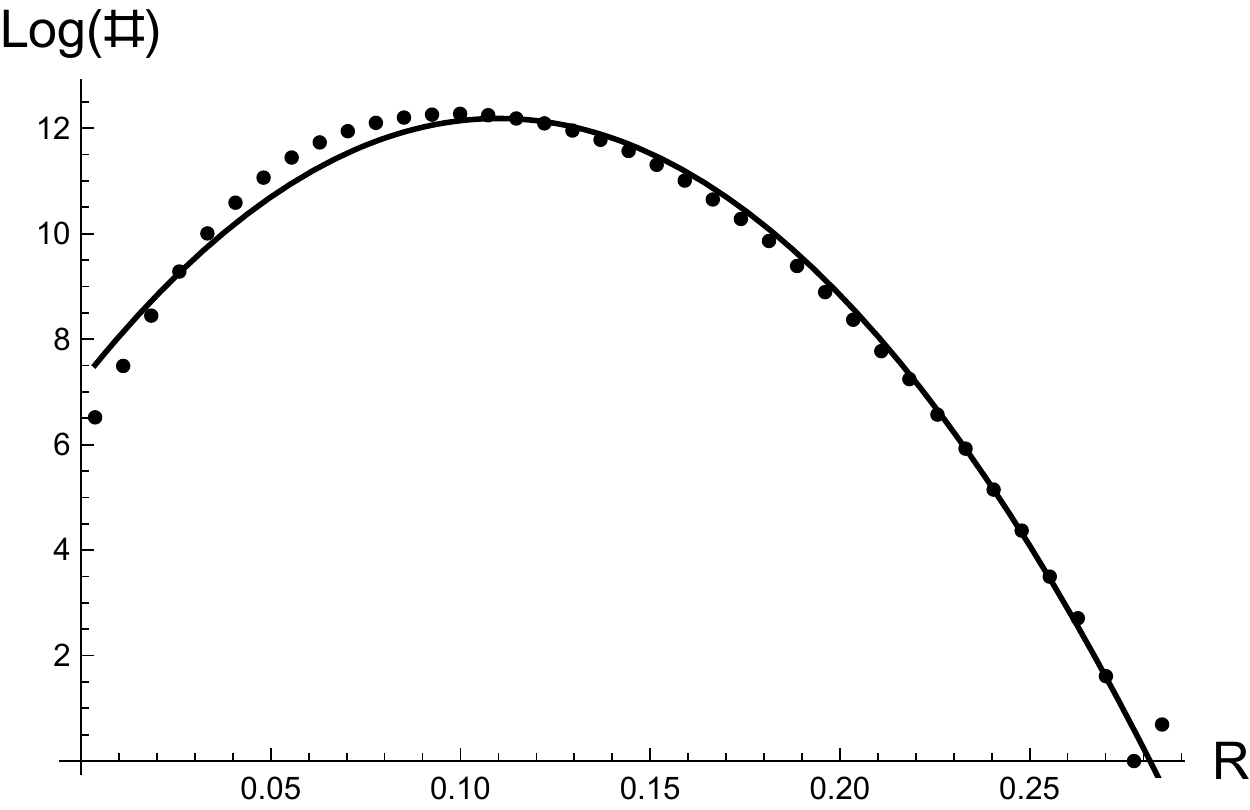}
   \includegraphics[width=3.2in]{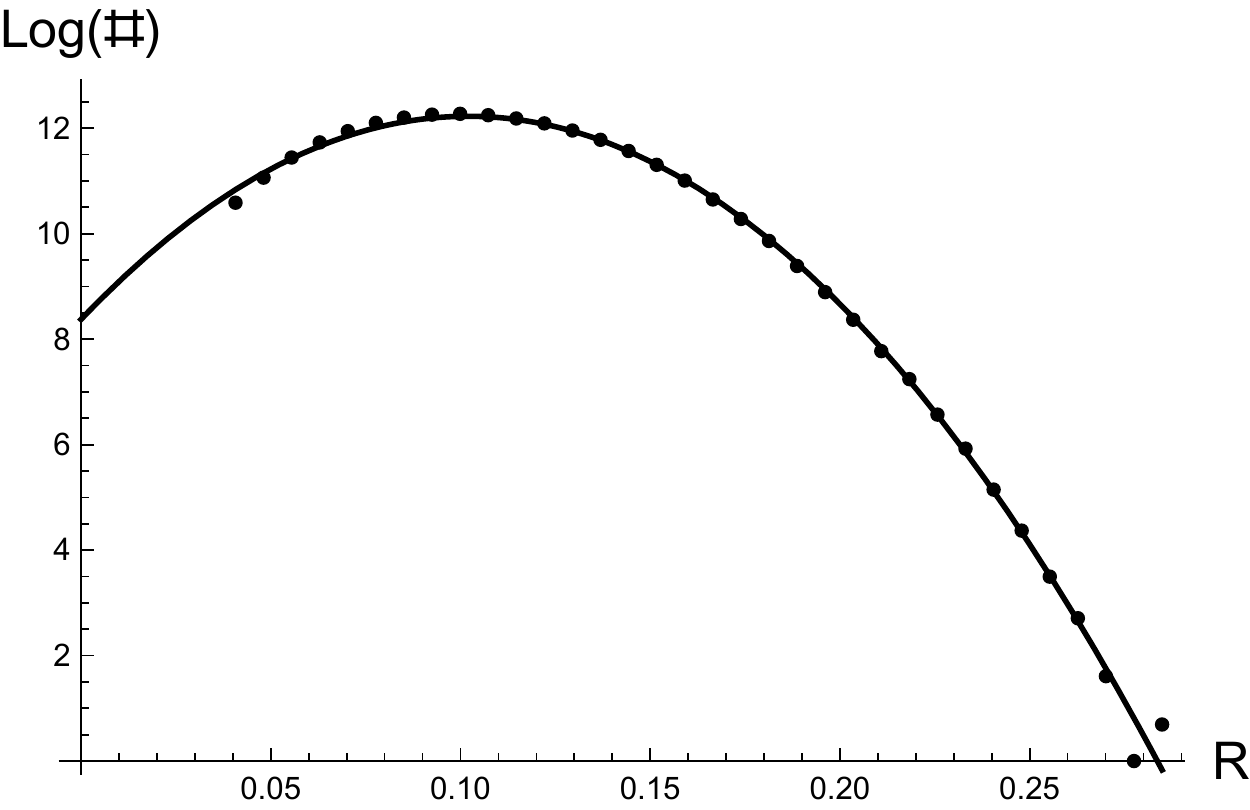}
    \caption{The distribution of $R$'s for 10 fields.  Here we used 160 realizations of the potential and sampled 20000 stationary points per realization.
        In the left graph we fit with the whole range of $R$'s and in the second we eliminated five points on the low end.  The latter distribution is a very good Gaussian.} 
       % We only had 160 realization of potentials and this may be due to this smaller number and also some bias against finding the very low minima.}
   \label{fig:Nf10}
\end{figure}

Like in all Monte-Carlo simulations, there is a danger of bias in the statistics. If the basins of attraction for different minima have significantly different sizes, we will have a biased sample. However, we have a guide: as we see in the smaller $N$ case, where we know the exact statistics, the distribution of $R$'s is a Gaussian. So, if after this sampling we find a Gaussian distribution for larger values of $N$, it may be a good indication that we are on the right track. We obtained the data for $N_c=30$ and $n_{\rm max}=10$ for different $N$'s. The results for all cases are very well fitted by Gaussians, except at very small values of $R$.  This may be due to  large walls surrounding very deep minima which makes them less accessible to the equations solver.  If the data at small $R$ did not fit well, we discarded the first few points, as illustrated in Fig.\ref{fig:Nf10}.

%For the largest number of fields, $N=10$, where we had a sample of 160 potential and we found 20000 stationary points per realization, we see that the low tail of $R$ does not fit well with a Gaussian, but the rest does. This may be due to  large walls surrounding very deep minima which makes them less accessible to the equations solver. These results are shown in Fig.\ref{fig:Nf10}. If the data at small $R$ did not fit well, we discarded the first few. 

For the equation solver, we used a Mathematica package developed by Ken Olum which uses a Powell Hybrid method for finding the roots of a given system of equation. This method,  which is a hybrid of gradient and Newton methods, has advantages over both of them: unlike the gradient methods it converges fast near the root and unlike the Newton, it does not take very large steps that may skip some roots.

\section{Numerical solution for bounces}\label{numericalBounce}

%We calculated bounce solutions which carry the tunneling according the shooting prescription given by Coleman in [?]. 
The bounce is an $O(4)$-symmetric solution of Euclidean field equations. 
As we outlined in Section IV, we reduced the  problem of tunneling in a multi-field landscape to a one-field problem. The bounce equation is given by
\bel{bounceEq}
	\phi''(r) + \frac3r \phi'(r)= \frac{dV}{d\phi}~,
\eeq
with boundary conditions 
\bel{boundary}
	\phi'(0)=0~, \qquad \lim_{\rho \rightarrow \infty}\phi(\rho) = \phi_{i}~,
\eeq
where $\phi_i$ is the value of $\phi$ in the false vacuum, prior to tunneling.
Equation \eqref{bounceEq} is a second order differential equation and needs two boundary conditions to specify a solution.
 %for numerical approaches. 
 We have to guess the value $\phi(0)$ in such a way  that after integrating to a large value of $r$ we reach the false vacuum. 
We wrote a shooting code in Mathematica to bracket the value of the field at the center of the bubble $\phi(0)$. We choose a value of $\phi(0)$ and integrate the bounce equation numerically until one of the following criteria is met:
\begin{enumerate}
	\item $\phi'$ vanishes. 
	\item $\phi$ passes the false vacuum.
\end{enumerate} 
The former  corresponds to overshoot and the latter to undershoot, and we can bracket the solution this way. However, there are many subtitles involved which we describe briefly here.

\subsection{Analytic evolution}
%Most of the times 
$\phi(0)$ is often too close to the true vacuum. Usually, if the solution resembles a thin-wall solution of radius $R$, we have $\phi(0)-\phi_{\rm tv} \sim e^{-m R}$ where $m$ is some mass scale of the problem. It is commonplace to encounter cases where $m R$ is of order of several hundreds. In these cases the machine precision is not enough for finding the solution and numerical integration invokes large errors. To overcome this problem, we use an analytic approximation until field moves away a small amount from the true vacuum. Near the true vacuum we can approximate the potential as  
\bel{UApp}
	V=V_{\rm tv}+\frac12 B \Phi^2~,
\eeq
where $\Phi=\phi-\phi_{\rm fv}$.
This simplifies the equations \eqref{bounceEq}
\bel{positive1}
 	\Phi''+ \frac3r\Phi'= B \Phi~.
\eeq
The solution to this equation with the appropriate boundary conditions can be written in terms of Bessel functions of type I. 
\beq
	\Phi(r) = 2 \Phi(0) \frac{I_1\left(\sqrt{B} r\right)}{\sqrt{B}r}~.
\eeq

\subsection{Non-uniqueness of solution}
\begin{figure}[htbp] %  figure placement: here, top, bottom, or page
   \centering
    \includegraphics[width=3in]{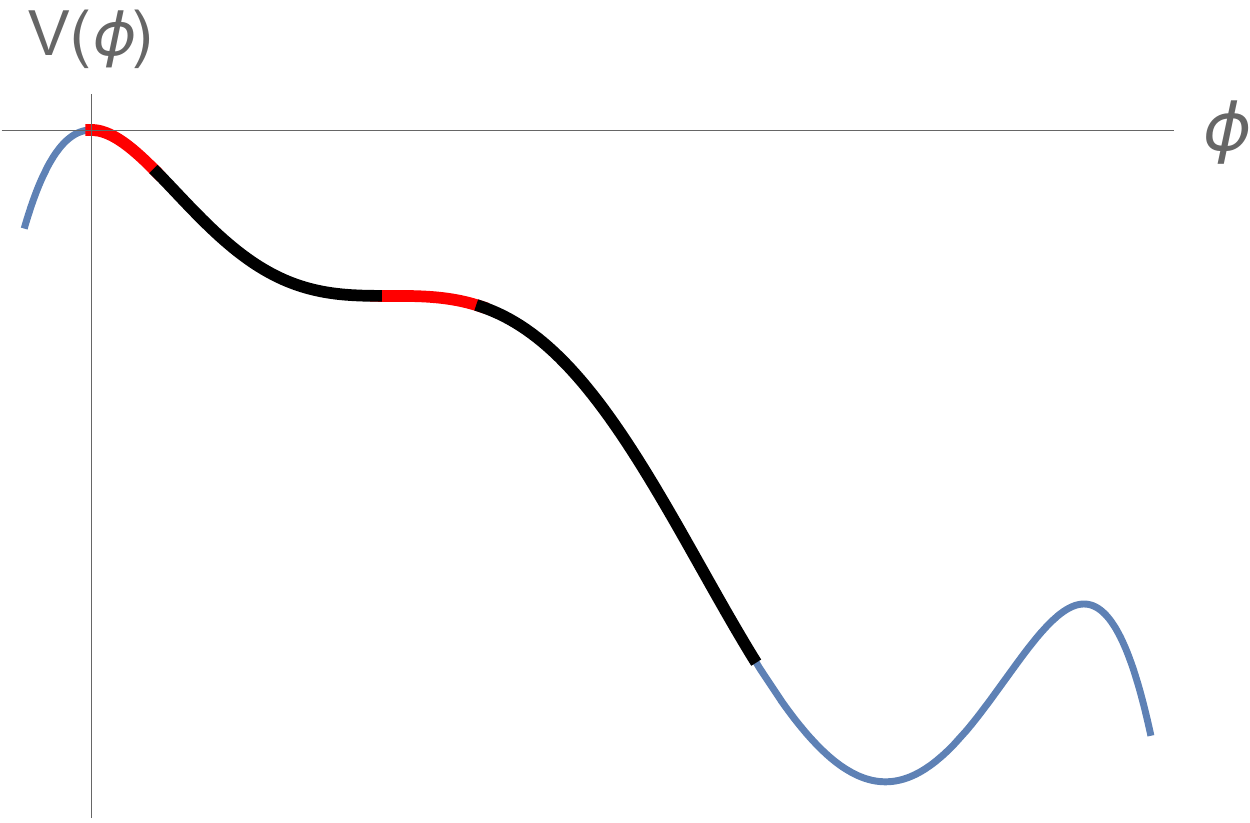} 
   \caption{A (inverted) potential which can possess more than one bounce solution. Red regions denote overshoot and black regions undershoot. At the boundary of these regions we can find Coleman bounces.}
   \label{fig:invertedPot}
\end{figure}

Solving Eq.~\eqref{bounceEq} is the same as evolving a field in the upside-down potentia. For a potential (already inverted to make it easier to see) shown in Fig.\ref{fig:invertedPot}, there can be up to three bounce solutions, and of course we should only use the one with the lowest action. These happen very rarely and therefore we did not look for multiple solutions. For each potential we tried to only find one solution. \footnote{We are thankful to Ken Olum for pointing out this possibility.} 

\subsection{Thin-wall solutions}
For the cases where $\phi(0)-\phi_{\rm tv} < 10^{-200} (\phi_{\rm fv} - \phi_{\rm tv})$, we did not solve the bounce equations. Instead, in these cases we directly used  the thin-wall approximation for calculation of the action. 

%\subsection{Action values}
%The tunneling action is the result of subtracting the action of the false vacuum from the bounce. Usually, both of these numbers can be large and the difference between them can be five order of magnitudes smaller than the individual values. This requires the code to be able to find the bounce action with very high precision. By patching different analytic and numerical methods we could achieve this accuracy.  Checking with few cases by hand, the code proved capable of generating the right numbers. 
\subsection{Larger number of fields}
To calculate the tunneling action we need to look at all possible channels of tunneling from a given vacuum to nearby vacua through all stationary points around it. Because the tunneling is dominated by the bounce with the least action, we can only calculate it if we know all the minima and stationary points around that vacuum. This is the bottleneck for the stability calculation. For the range of parameters we choose, we can only find all the minima and stationary points for $N\le4$ and for this reason we did not study the stability for larger number of fields. 

%\clearpage
%\setcounter{page}{1}

\end{document}